\documentclass[lettersize,journal]{IEEEtran}
\usepackage{amsmath,amsfonts}
\usepackage{algorithmic}
\usepackage{algorithm}
\usepackage{array}
\usepackage[caption=false,font=scriptsize,labelfont=rm,textfont=rm]{subfig}
\usepackage{textcomp}
\usepackage{stfloats}
\usepackage{url}
\usepackage{verbatim}
\usepackage{graphicx}
\usepackage{cite}
\usepackage{bm}			
\usepackage{amssymb}		
\usepackage{mathrsfs}			
\usepackage[colorlinks=true,citecolor=blue,linkcolor=black,urlcolor=blue]{hyperref}

\begin{document}

\title{ Movable Antenna Enhanced DF and AF Relaying Systems: Performance Analysis and Optimization }

\author{Nianzu Li,~Weidong Mei,~\IEEEmembership{Member,~IEEE},~Peiran Wu,~\IEEEmembership{Member,~IEEE},~Boyu Ning,~\IEEEmembership{Member,~IEEE},\\~and Lipeng Zhu,~\IEEEmembership{Member,~IEEE}
\thanks{Part of this paper has been presented in IEEE Global Communications Conference 2024 Workshops \cite{ref8}.}
\thanks{N. Li and P. Wu are with School of Electronics and Information Technology, Sun Yat-sen University, Guangzhou 510006, China (e-mail: linz5@mail2.sysu.edu.cn; wupr3@mail.sysu.edu.cn).}
\thanks{W. Mei and B. Ning are with National Key Laboratory
	of Wireless Communications, University of Electronic
	Science and Technology of China, Chengdu 611731, China (e-mail: wmei@uestc.edu.cn;  boydning@outlook.com).}
\thanks{L. Zhu is with the Department of Electrical and Computer
	Engineering, National University of Singapore, Singapore 117583 (e-mail: zhulp@nus.edu.sg).}
}        



\maketitle

\begin{abstract}
Movable antenna (MA) has been deemed as a promising technology to flexibly reconfigure wireless channels by adjusting the antenna positions in a given local region, which provides new
degrees of freedom (DoFs) to enhance wireless communication performance. In this paper, we investigate the application of the MA technology in both decode-and-forward (DF) and amplify-and-forward (AF) relaying systems, where a relay is equipped with multiple MAs to assist in the data transmission between two single-antenna nodes. For the DF relaying system, our objective is to maximize the achievable rate at the destination by jointly optimizing the  positions of the MAs in two stages for receiving signals from the source and transmitting signals to the destination, respectively. To drive essential insights, we first derive a closed-form upper bound on the maximum achievable rate of the DF relaying system. Then, a low-complexity algorithm based on projected gradient ascent (PGA) and alternating optimization (AO) is proposed to solve the antenna position optimization problem. For the AF relaying system, our objective is to maximize the achievable rate by jointly optimizing the two-stage MA positions as well as the AF beamforming matrix at the relay, which results in a more challenging optimization problem due to the intricate coupling variables. To tackle this challenge, we first reveal the hidden separability among the antenna position optimization in the two stages and the beamforming optimization. Based on such separability, we derive a closed-form upper bound on the maximum achievable rate of the AF relaying system and propose a low-complexity algorithm to obtain a high-quality suboptimal solution to the considered problem. Simulation results validate the efficacy of our theoretical analysis and demonstrate the superiority of the MA-enhanced relaying systems to the conventional relaying systems with fixed-position antennas (FPAs) and other benchmark schemes.
\end{abstract}


\section{Introduction}
\IEEEPARstart{D}{riven} by the increasing demand of larger capacity and higher reliability in the next-generation wireless system, multiple-input multiple-output (MIMO) and massive MIMO technologies have been widely promoted and investigated in both academia and industry\cite{ref39,ref40}. In particular, due to the limited spectrum resources, conventional single-antenna or single-input single-output (SISO) technology generally cannot support massive and heterogeneous data transmission. Instead, MIMO communication technology can significantly enhance the channel capacity and spectral efficiency by exploiting the pronounced spatial multiplexing gain, which allows parallel transmission of multiple data streams in the same time-frequency resource block\cite{ref41}. With the advancement of communication systems towards higher frequency bands, such as millimeter-wave (mmWave) and terahertz (THz) bands, a larger-scale antenna array, i.e., massive MIMO technology, has also been developed to provide even more significant spatial multiplexing and beamforming gains, which can compensate for more severe propagation loss over high frequencies\cite{ref42,ref43,ref44,ref59}. 

However, conventional MIMO and massive MIMO technologies may lead to high energy consumption and hardware cost in practice due to the growing number of radio frequency (RF) chains, especially in the high frequency band. To overcome this limitation, the antenna selection (AS) technology has been proposed as a viable approach to achieve favorable wireless MIMO channels by using the RF switches to select a subset of antennas with good channel conditions from a given set of antennas, which can alleviate the requirement on the number of RF chains and become attractive for massive MIMO systems\cite{ref45,ref46,ref47}. Another approach is by designing the hybrid beamforming antenna architecture, where the overall beamformer comprises a low-dimensional digital beamformer followed by some RF beamformers implemented using the analog phase shifters\cite{ref48,ref49}. This technology can achieve a close
performance to the fully digital scheme with much fewer RF chains. Nonetheless, the hardware cost required for a large number of antennas or phase shifters in high-frequency MIMO systems still remains a critical issue. Moreover, by only relying on the traditional fixed-position antennas (FPAs), the spatial variation of the wireless channels cannot be fully exploited in the given transmit and receive regions, which may result in suboptimal spatial diversity/multiplexing performance\cite{ref1}.

Recently, movable antenna (MA) and fluid antenna system (FAS) have been proposed as potential technologies to resolve the above issues, owing to their capability of reconfiguring the wireless channels via antenna movement\cite{ref1,ref24,ref16,ref35}. Specifically, MAs are connected to the RF chains via flexible wires, such as coaxial cables, and their positions can be mechanically adjusted within a given continuous spatial region with the aid of driving components, such as stepper motors and servos. Different from conventional FPAs that undergo random wireless channels or may suffer from deep fading, the MA system can dynamically move the antennas to the positions with more favorable channel conditions, thus providing additional degrees of freedom (DoFs) in the spatial domain to enhance the wireless communication performance without the need for increasing the number of antennas and RF chains. Similarly, the FAS uses conductive fluids or electronically reconfigurable pixels as electromagnetic radiators, which can freely select the favorable antenna positions from a given set of candidate ports for reconfiguring the wireless channels\cite{ref24}. Besides, a novel six-dimensional movable antenna (6DMA) architecture was developed most recently, which consists of distributed antenna surfaces with the capability of independently adjusting the three-dimensional (3D) positions as well as the 3D rotations in a given local space\cite{ref64,ref50}. This 6DMA system provides full DoFs for reconfiguring the wireless channels and thus can reap the spatial diversity multiplexing more efficiently. Due to the promising benefits of MAs/FASs, they can find a wide range of applications in future wireless networks, such as machine-type communications, cellular communications, and satellite communications\cite{ref1}.

 Motivated by the above, existing works have investigated the antenna position optimization problem for the MA/FAS-assisted systems and demonstrated their superior communication performance to conventional FPA-based systems under different scenarios\cite{ref3,ref27,ref4,ref62,ref5,ref69,ref55,ref60,ref70,ref10,ref12,ref61,ref2,ref54,ref67,ref7,ref56,ref29,ref58,ref28,ref68,ref63}. The authors in \cite{ref3} developed a field response-based channel model for MAs and analyzed the maximum signal-to-noise ratio (SNR) gain under both deterministic and stochastic channels. In \cite{ref27}, the authors developed a graph-based algorithm to obtain the optimal MA positions in a multiple-input single-output (MISO) communication system by discretizing the transmit region into a multitude of sampling points. In \cite{ref4}, the authors studied a MIMO communication system assisted by both transmit and receive MAs. In \cite{ref62,ref5,ref69,ref55,ref60,ref70}, the authors studied the utility of MAs/FASs for enhancing the physical-layer security. Moreover, combining MAs/FASs with various multiple access technologies has also become a hot topic. For example, the authors in \cite{ref10,ref12,ref61} investigated their applications in space-division multiple access (SDMA) systems and the authors in \cite{ref2,ref54,ref67} extended them into non-orthogonal multiple access (NOMA) systems.  Furthermore, MAs/FASs have also been applied to other system setups, such as over-the-air computation\cite{ref7}, multicast communication\cite{ref56}, cognitive radio\cite{ref29}, flexible beamforming\cite{ref58,ref28,ref68}, and index-modulated transmission\cite{ref63}.
 
\begin{figure}[t]
	\centering
	\includegraphics[width=0.49\textwidth]{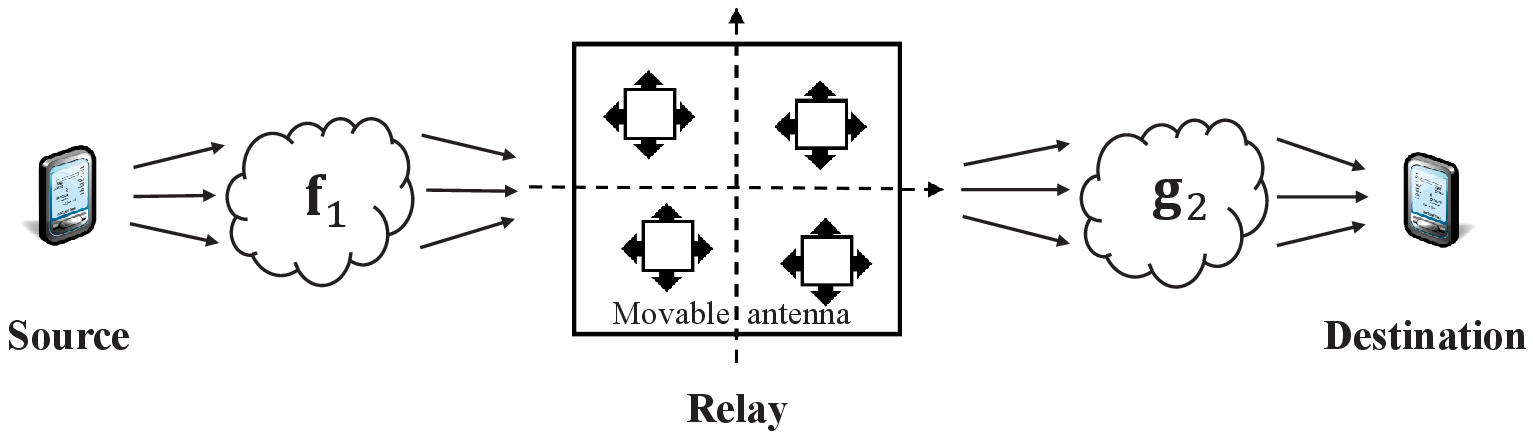}
	\caption{MA-enhanced relaying system.}
	\label{system_model}
\end{figure}

For relaying systems, the limitation of conventional FPA technologies becomes more pronounced, as the overall transmission rate may be restrained by the deep fading channels in either the source-relay link or the relay-destination link. In this context, MAs can efficiently improve the channel conditions for both links by adjusting the antenna positions. However, to the best of our knowledge, MAs have not been applied to a relaying system so far in the literature. To fill in this gap, we advance their application in this paper for both decode-and-forward (DF) and amplify-and-forward (AF) relaying systems. The main contributions of our work are summarized as follows:
\begin{itemize}
	\item We consider an MA-enhanced relaying system, where a single-antenna source aims to communicate with a single-antenna destination with the aid of a multi-MA relay, as shown in Fig \ref{system_model}. Unlike most related works focusing on one-time position adjustment, the MA positions in the considered system can be adjusted in two stages for receiving signals from the source and transmitting signals to the destination, respectively. Then, we investigate the performance upper bound and optimization of this system under both DF relaying and AF relaying architectures.
	
	\item First, for the DF relaying system, we derive an upper bound on the maximum achievable rate at the destination in closed form by assuming the size of the MA moving region is arbitrarily large, which unveils the significant performance gains of MAs over conventional FPAs. Then, a low-complexity algorithm based on projected gradient ascent (PGA) and alternating optimization (AO) methods is proposed to jointly optimize the two-stage MA positions within a finite moving region.
	
	\item Next, for the AF relaying system, the problem becomes more challenging to be optimally solved since the two-stage MA positions and the AF beamforming matrix are highly coupled with each other in the achievable rate at the destination. To tackle this issue, we reveal the hidden separability among these optimization variables, based on which an upper bound on the maximum achievable rate is derived in closed form to drive essential insights. Furthermore, by exploiting the separability, we properly modify the proposed algorithm for the DF relaying system to obtain a high-quality suboptimal solution for the AF relaying system.
	
	\item Finally, numerical results are provided to evaluate the efficacy of our analysis and the performance of our proposed algorithms for the MA-enhanced relaying systems. It is shown that our theoretical analysis is well-consistent with the numerical results. Besides, results also show that our proposed schemes can significantly improve the achievable rate compared with the conventional FPA-based relaying schemes and other benchmark schemes. Moreover, our proposed algorithms can achieve a small performance gap with the derived analytical upper bounds, which thus provide a viable and attractive
	solution for enhancing the relaying performance in practice.
\end{itemize}

The rest of this paper is organized as follows. In Section II, we present the system model of the considered MA-enhanced relaying systems. In Section III, we formulate two optimization problems to maximize the achievable rates for both DF and AF relaying systems. In Sections IV and V, we present the performance analysis and optimization algorithms for the MA-enhanced DF relaying and AF relaying systems, respectively. Simulation results are provided in Section VI and conclusions are finally drawn in Section VII.

\textit{Notations}: Boldface lowercase and uppercase letters denote vectors and matrices, respectively. $(\cdot)^\mathrm{T}$, $(\cdot)^{*}$, and $(\cdot)^\mathrm{H}$ denote transpose, conjugate, and conjugate transpose, respectively. $\mathcal{CN}(0,\mathbf{\Sigma})$ denotes the circularly symmetric complex Gaussian (CSCG) distribution with zero mean and covariance matrix $\mathbf{\Sigma}$. $\|\mathbf{A}\|_{\ell}$, $\|\mathbf{A}\|$, $\mathrm{rank}(\mathbf{A})$, and $\mathrm{Tr}(\mathbf{A})$ denote the $\ell$-norm, Frobenius norm, rank, and trace of the matrix $\mathbf{A}$, respectively. $\mathrm{vec}(\mathbf{A})$ is the vectorization of the matrix $\mathbf{A}$. $\otimes$ is the Kronecker product. $\mathbb{E}\left\{\cdot\right\}$ is the expectation of a random variable. $n!$ is the factorial of a non-negative integer $n$, defined as the product of all positive integers from $1$ to $n$. $(2n-1)!!$ is the double factorial of an odd integer $2n-1$, defined as the product of all odd integers from $1$ to $2n-1$.

\section{System Model}
\begin{figure}[t]
	\centering
	\includegraphics[width=0.45\textwidth]{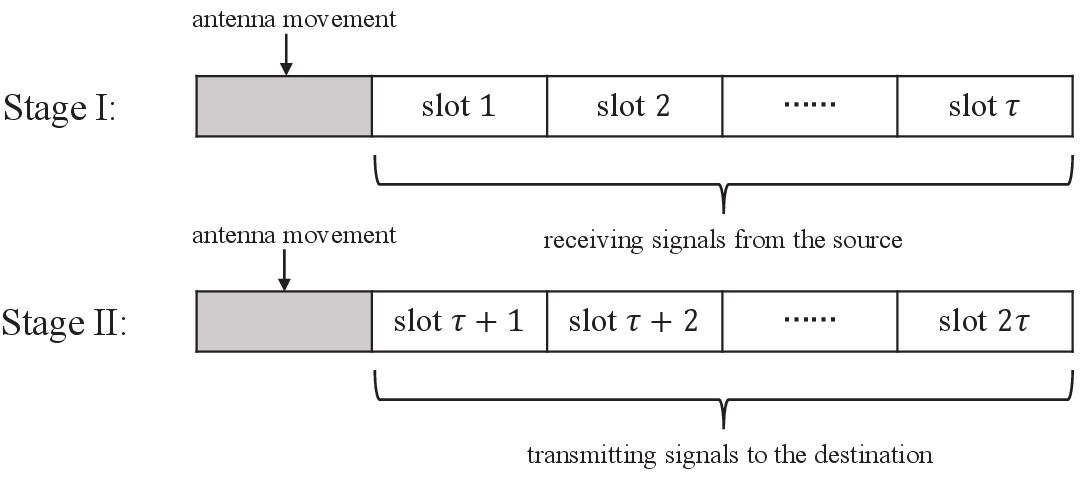}
	\caption{Two stages of the information reception and transmission at the relay.}
	\label{data_transmission}
\end{figure}
As shown in Fig. \ref{system_model}, we consider an MA-enhanced relaying system, consisting of a single-antenna source, a multi-antenna relay, and a single-antenna destination. The source and the destination are both equipped with an FPA, and the direct link between them is assumed to be severely blocked due to the dense obstacles in the environment\footnote{This can practically occur in device-to-device communications, where two remote sensors aim to communicate with each other via multi-hop data transmission with the aid of the relay.}. Therefore, a relay equipped with $N$ MAs is deployed to assist in their data transmission. We assume that each MA at the relay can be flexibly moved within a
local two-dimensional (2D) region, denoted as $\mathcal{C}_{\mathrm{r}}$. Without loss of generality, we consider a square moving area of size $A \times A$, with $A$ denoting the side length of each dimension. Besides, we assume that the relay operates in the half-duplex mode, such that the data transmission can be divided into two stages\cite{ref36}, i.e., Stage I and Stage II, each containing $\tau$ time slots.  As shown in Fig. \ref{data_transmission}, the relay first receives the signals from the source in Stage I, and then forwards them to the destination in Stage II. 

In order to facilitate the information reception/transmission from/to the source/destination, the positions of the MAs at the relay are adjusted at the beginning of the two stages. 
Denote the collections of all MAs' positions in Stage I and Stage II as $\tilde{\mathbf{r}}=[\mathbf{r}_1,\cdots,\mathbf{r}_N]\in\mathbb{C}^{2\times N}$ and $\tilde{\mathbf{t}}=[\mathbf{t}_1,\cdots,\mathbf{t}_N]\in\mathbb{C}^{2\times N}$, respectively, where $\mathbf{r}_n=[x_{\mathrm{r},n},y_{\mathrm{r},n}]^\mathrm{T}\in\mathcal{C}_{\mathrm{r}},\mathbf{t}_n=[x_{\mathrm{t},n},y_{\mathrm{t},n}]^\mathrm{T}\in\mathcal{C}_{\mathrm{r}},1\leq n \leq N,$ represent the Cartesian coordinates of the $n$-th MA. As a result, the channel vector from the source/relay to the relay/destination can be denoted as $\mathbf{h}_1(\tilde{\mathbf{r}})\in\mathbb{C}^{N\times1}$ and $\mathbf{h}_2(\tilde{\mathbf{t}})\in\mathbb{C}^{N\times1}$, respectively, each of which is a function of the MA positions at the relay. 

In this paper, we consider narrow-band quasi-static channels, where all the channel vectors involved remain approximately constant during each fading block\footnote{We assume that the antenna movement delay is much smaller than the channel coherence time, which usually holds in various slow-varying scenarios, e.g., smart homes and smart factories. On the other hand, we can also reduce the movement delay by resorting to more advanced electronically-driven solutions to realize equivalent antenna repositioning \cite{ref35}. }. 
To model the channel in terms of the MA positions, we apply the field-response channel model proposed in \cite{ref3}. Let $L_\mathrm{r}$ denote the number of receive channel paths at the relay from the source. Then, denote $\theta_{1,\ell}$ and $\phi_{1,\ell}$ as the corresponding elevation and azimuth angles of arrival (AoAs) of the $\ell$-th $(1\leq \ell \leq L_\mathrm{r})$ receive channel path. Thus, the receive field-response vector (FRV) for the multiple channel paths between the source and the $n$-th MA at the relay is given by
\begin{equation}
	\label{eq1}
	\mathbf{f}_1(\mathbf{r}_n)=\left[e^{\mathrm{j}\frac{2\pi}{\lambda}\omega_{1,1}(\mathbf{r}_n)},\cdots,e^{\mathrm{j}\frac{2\pi}{\lambda}\omega_{1,L_\mathrm{r}}(\mathbf{r}_n)}\right]^\mathrm{T},\forall n,
\end{equation}
where $\omega_{1,\ell}(\mathbf{r}_n)=x_{\mathrm{r},n}\sin\theta_{1,\ell}\cos\phi_{1,\ell}+y_{\mathrm{r},n}\cos\theta_{1,\ell}$ characterizes the phase difference between the $n$-th MA at the relay and its reference point (denoted as $\mathbf{o}=[0,0]^\mathrm{T}$), and $\lambda$ denotes the carrier wavelength. As a result, in Stage I, the channel from the source to the relay can be represented as
\begin{equation}
	\label{eq2}
	\mathbf{h}_1(\tilde{\mathbf{r}})=\mathbf{F}_1^\mathrm{H}(\tilde{\mathbf{r}})\mathbf{g}_1=\left[\mathbf{f}_1(\mathbf{r}_1),\cdots,\mathbf{f}_1(\mathbf{r}_N)\right]^\mathrm{H}\mathbf{g}_1,
\end{equation}
where $\mathbf{g}_1=[g_{1,1},\cdots,g_{1,L_\mathrm{r}}]^{\mathrm{T}}\in\mathbb{C}^{L_\mathrm{r}\times1}$ is the path-response vector (PRV) from the source to the relay with independent and identically distributed (i.i.d.) CSCG random elements, i.e., $g_{1,\ell}\sim\mathcal{CN}(0,\rho_1^2/L_\mathrm{r}),1\leq \ell \leq L_\mathrm{r}$, representing the multi-path channel response coefficients from the source to the reference point at the relay.
Similarly, let $L_\mathrm{t}$ denote the total number of transmit channel paths from the relay to the destination. Then, denote $\theta_{2,\ell}$ and $\phi_{2,\ell}$ as the corresponding elevation and azimuth angles of departure (AoDs) of the $\ell$-th $(1 \leq \ell \leq L_\mathrm{t})$ transmit channel path. Thereby, the transmit FRV between the $n$-th MA at the relay and the destination is given by
\begin{equation}
	\label{eq3}
	\mathbf{g}_2(\mathbf{t}_n)=\left[e^{\mathrm{j}\frac{2\pi}{\lambda}\omega_{2,1}(\mathbf{t}_n)},\cdots,e^{\mathrm{j}\frac{2\pi}{\lambda}\omega_{2,L_\mathrm{t}}(\mathbf{t}_n)}\right]^\mathrm{T},\forall n,
\end{equation}
where $\omega_{2,\ell}(\mathbf{t}_n)=x_{\mathrm{t},n}\sin\theta_{2,\ell}\cos\phi_{2,\ell}+y_{\mathrm{t},n}\cos\theta_{2,\ell}$. Accordingly, in Stage II, the channel from the relay to the destination can be represented as
\begin{equation}
	\label{eq4}
	\mathbf{h}_2(\tilde{\mathbf{t}})=\mathbf{G}_2^\mathrm{H}(\tilde{\mathbf{t}})\mathbf{f}_2=\left[\mathbf{g}_2(\mathbf{t}_1),\cdots,\mathbf{g}_2(\mathbf{t}_N)\right]^\mathrm{H}\mathbf{f}_2,
\end{equation}
where $\mathbf{f}_2=[f_{2,1},\cdots,f_{2,L_\mathrm{t}}]^{\mathrm{T}}\in\mathbb{C}^{L_\mathrm{t}\times1}$ is the PRV from the relay to the destination with i.i.d. CSCG random elements, i.e., $f_{2,\ell}\sim\mathcal{CN}(0,\rho_2^2/L_\mathrm{t}),1\leq \ell \leq L_\mathrm{t}$, representing the multi-path channel response coefficients from the reference point at the relay to the destination.

In this paper, we consider two relaying strategies at the relay, i.e., DF and AF. The associated signal models are presented as below.

\subsection{MA-Enhanced DF Relaying System}
For the considered MA-enhanced DF relaying system, the source first transmits its information to the relay in Stage I and the received signal at the relay can be expressed by  
\begin{equation}
	\label{eq5}
	\mathbf{y}_\mathrm{r}(t)=\sqrt{P_\mathrm{s}}\mathbf{h}_1(\tilde{\mathbf{r}}) s(t)+\mathbf{n}_\mathrm{r}(t),\forall t\in\{1,\cdots,\tau\},
\end{equation}
where $P_\mathrm{s}$ is the source transmit power, $s(t)$ is the transmitted symbol from the source at the $t$-th time slot with unit power and $\mathbf{n}_\mathrm{r}(t)\sim\mathcal{CN}(\mathbf{0},\sigma_\mathrm{r}^2\mathbf{I}_N)$ is the additive white Gaussian noise (AWGN) at the relay. Since the DF strategy is employed, the relay will decode the source data from the received signals and then transmit the regenerated symbol to the destination in Stage II. Thus, the received signal at the destination can be expressed by
\begin{equation}
	\label{eq6}
	y_\mathrm{d}(t)=\mathbf{h}_2^\mathrm{H}(\tilde{\mathbf{t}})\mathbf{w}_\mathrm{r}\hat{s}(t)+n_\mathrm{d}(t),~\forall t \in \{\tau+1,\cdots,2\tau\},
\end{equation}
where $\mathbf{w}_\mathrm{r}$ is the beamforming vector at the relay, $\hat{s}(t)$ is the regenerated symbol with unit power, and $n_\mathrm{d}(t)$ is the AWGN at the destination, with variance $\sigma_\mathrm{d}^2$. Thereby, the total transmit power of the relay is $\|\mathbf{w}_\mathrm{r}\|^2$, which is constrained by the total relay power budget, denoted as $P_{\mathrm{r}}$, i.e., $\|\mathbf{w}_\mathrm{r}\|^2\leq P_{\mathrm{r}}$. 

In particular, to maximize the received signal-to-noise ratio (SNR) at the relay/destination, the maximum ratio combining (MRC) and maximum ratio transmission (MRT) are employed at the relay to decode the information from the source in Stage I and transmit its regenerated version in Stage II, respectively. Therefore, the end-to-end achievable rate at the destination for the DF relaying system is given by
\begin{equation}
	\label{eq7}
	R_{\mathrm{DF}}=\frac{1}{2}\log_2\left(1+\min\left\{\frac{P_\mathrm{s}\|\mathbf{h}_1(\tilde{\mathbf{r}})\|^2}{\sigma_\mathrm{r}^2},\frac{P_{\mathrm{r}}\|\mathbf{h}_2(\tilde{\mathbf{t}})\|^2}{\sigma_\mathrm{d}^2}\right\}\right), 
\end{equation}
where 1/2 is due to the half-duplex processing of the relay.

\subsection{MA-Enhanced AF Relaying System}
For the considered MA-enhanced AF relaying system, the relay will process the received signals from the source by a beamforming matrix $\mathbf{W}\in\mathbb{C}^{N\times N}$, and then forward them to the destination.  Specifically, in Stage I, the received signal at the relay can be expressed by
\begin{equation}
	\label{eq6}
	\mathbf{y}_\mathrm{r}(t)=\sqrt{P_\mathrm{s}}\mathbf{h}_1(\tilde{\mathbf{r}}) s(t)+\mathbf{n}_\mathrm{r}(t),\forall t\in\{1,\cdots,\tau\}.
\end{equation}
Then, in Stage II, the received signal at the destination is expressed by
\begin{align}
	y_\mathrm{d}(t)=\sqrt{P_\mathrm{s}}\mathbf{h}_2^{\mathrm{H}}(\tilde{\mathbf{t}})&\mathbf{W}\mathbf{h}_1(\tilde{\mathbf{r}})s(t-\tau)+\mathbf{h}_2^{\mathrm{H}}(\tilde{\mathbf{t}})\mathbf{W}\mathbf{n}_\mathrm{r}(t-\tau)\notag\\
	&+n_\mathrm{d}(t),~\forall t \in \{\tau+1,\cdots,2\tau\}.
\end{align}
Note that since the source signals are not decoded at the relay, noise amplification will also be introduced at the same time. As such, the end-to-end achievable rate at the destination for the AF relaying system is given by
\begin{equation}
	\label{eq22}
	R_{\mathrm{AF}}=\frac{1}{2}\log_2\left(1+\frac{P_\mathrm{s}\left|\mathbf{h}_2^\mathrm{H}(\tilde{\mathbf{t}})\mathbf{W}\mathbf{h}_1(\tilde{\mathbf{r}})\right|^2}{\sigma_\mathrm{r}^2\left\|\mathbf{h}_2^\mathrm{H}(\tilde{\mathbf{t}})\mathbf{W}\right\|^2+\sigma_\mathrm{d}^2}\right).
\end{equation}

\section{Problem Formulation}
In this paper, our objective is to maximize the achievable rate at the destination for both the MA-enhanced DF relaying and AF relaying systems. Specifically, for the considered DF relaying system, we aim to maximize (7) by jointly optimizing the two-stage MA positions, which thus yields the following optimization problem:
\begin{subequations}
	\begin{align}
		\text{(P1)} \quad \max_{\tilde{\mathbf{r}},\tilde{\mathbf{t}}}& \quad R_{\mathrm{DF}}\\
		\mathrm{s.t.}
		& \quad \mathbf{r}_n\in\mathcal{C}_\mathrm{r},~1\leq n\leq N,\\
		& \quad \mathbf{t}_n\in\mathcal{C}_\mathrm{r},~1\leq n\leq N,\\
		& \quad \|\mathbf{r}_m-\mathbf{r}_n\|_2\geq D,~\forall m\neq n,\\
		& \quad \|\mathbf{t}_m-\mathbf{t}_n\|_2\geq D,~\forall m\neq n,
	\end{align}
\end{subequations}
where constraints (11b) and (11c) confine the MAs moving in their feasible region; constraints (11d) and (11e) ensure that the minimum inter-MA distance $D$ is satisfied to avoid the mutual coupling between each two antennas. 

Besides, for the considered AF relaying system, our goal is to maximize (10) by jointly optimizing the two-stage MA positions as well as the beamforming matrix at the relay. Thus, the optimization problem can be formulated as
\begin{subequations}
	\begin{align}
		\text{(P2)} \quad \max_{\mathbf{W},\tilde{\mathbf{r}},\tilde{\mathbf{t}}}& \quad R_{\mathrm{AF}}\\
		\mathrm{s.t.}
		& \quad 
		P_\mathrm{s}\|\mathbf{W}\mathbf{h}_1(\tilde{\mathbf{r}})\|^2+\sigma_\mathrm{r}^2\|\mathbf{W}\|^2\leq P_{\mathrm{r}},\\
		& \quad \mathbf{r}_n\in\mathcal{C}_\mathrm{r},~1\leq n\leq N,\\
		& \quad \mathbf{t}_n\in\mathcal{C}_\mathrm{r},~1\leq n\leq N,\\
		& \quad \|\mathbf{r}_m-\mathbf{r}_n\|_2\geq D,~\forall m\neq n,\\
		& \quad \|\mathbf{t}_m-\mathbf{t}_n\|_2\geq D,~\forall m\neq n,
	\end{align}
\end{subequations}
where constraint (12b) indicates that the total transmit power at the relay can not exceed its maximum value $P_\mathrm{r}$. 

However, both (P1) and (P2) are difficult to be optimally solved due to the non-concave objective functions and the continuous nature of the MA positions. Besides, (P2) is even more challenging to solve than (P1) due to the intricate coupling between the MA positions and the AF beamforming matrix. In Sections IV and V, we will first analyze the performance upper bounds of the considered MA-enhanced relaying systems and propose low-complexity algorithms to solve (P1) and (P2).

\section{Performance Analysis and Proposed\\~Solution for DF Relaying}
In this section, we focus on DF relaying and solving (P1). First, we analyze the upper bound on $R_{\mathrm{DF}}$ by assuming an infinitely large MA moving region. Then, we propose a low-complexity algorithm to solve (P1) with a finite MA moving region.

\subsection{Performance Analysis}

From (P1), it is observed that the two optimization variables, i.e.,  $\tilde{\mathbf{r}}$ and $\tilde{\mathbf{t}}$, are mutually independent. Thereby, the joint optimization
problem can be equivalently decomposed into two separate subproblems:
\begin{align}
		&\text{(P1.1)}~\max_{\tilde{\mathbf{r}}} \quad \|\mathbf{h}_1(\tilde{\mathbf{r}})\|^2 \quad \mathrm{s.t.} \quad \text{(11b), (11d)},\\
		&\text{(P1.2)}~\max_{\tilde{\mathbf{t}}} \quad \|\mathbf{h}_2(\tilde{\mathbf{t}})\|^2 \quad \mathrm{s.t.} \quad \text{(11c), (11e)},
\end{align}
which maximize the source-relay channel gain and the relay-destination channel gain, respectively. Accordingly, if the channel gains in both receiving and transmitting phases are equal to their respective maximum values, i.e., the amplitude of each element in $\mathbf{h}_1(\tilde{\mathbf{r}})\triangleq\left[h_{1,1}(\mathbf{r}_1),\cdots,h_{1,N}(\mathbf{r}_N)\right]^{\mathrm{T}}$, and $\mathbf{h}_2(\tilde{\mathbf{t}})\triangleq\left[h_{2,1}(\mathbf{t}_1),\cdots,h_{2,N}(\mathbf{t}_N)\right]^{\mathrm{T}}$ is equal to its maximum value, the achievable rate in (7) can achieve an upper bound. This indicates that to achieve the maximum achievable rate, all MAs should be moved to the positions with the highest channel  gains subjected to the inter-MA distance constraint. In addition, it is also noted that the source-relay and the relay-destination receive SNR have a comparable effect on the achievable rate in (7), which implies that if  the channel condition between the source and the relay is similar to that between the relay and the destination, the solutions to these two subproblems, i.e., (P1.1) and (P1.2), should be similar as well. In such case, the positions of the MAs only need to be slightly adjusted to cater to both the information reception and transmission without significant performance loss. 

To characterize the performance upper bound of the DF relaying system, we present the following lemma.

\textit{\textbf{Lemma 1:} If the size of the MA moving region at the relay is sufficiently large, a tight upper bound on the maximum channel gain between the source/relay and the relay/destination can be achieved as}
\begin{subequations}
	\begin{align}
		&\|\mathbf{h}_1(\tilde{\mathbf{r}})\|^2=\sum_{n=1}^{N}\left|\mathbf{f}_1^{\mathrm{H}}(\mathbf{r}_n)\mathbf{g}_1\right|^2\leq N\|\mathbf{g}_1\|_1=N\left(\sum_{\ell=1}^{L_\mathrm{r}}\left|g_{1,\ell}\right|\right)^2,\\
		&\|\mathbf{h}_2(\tilde{\mathbf{t}})\|^2=\sum_{n=1}^{N}\left|\mathbf{g}_2^{\mathrm{H}}(\mathbf{t}_n)\mathbf{f}_2\right|^2\leq N\|\mathbf{f}_2\|_1 =N\left(\sum_{\ell=1}^{L_\mathrm{r}}\left|f_{2,\ell}\right|\right)^2.
	\end{align}
\end{subequations}

\textit{Proof: The proof is similar to \cite{ref3} and thus omitted in this paper for brevity. $\hfill\square$}

Lemma 1 characterizes an upper bound on the maximum channel gain achieved by MAs if the arbitrarily large region (ALR) assumption is satisfied. Note that this upper bound can be approached by aligning the phase of the FRV to that of the PRV via antenna position optimization. Thus, by substituting (15) into (7), we can derive an upper bound on $R_{\mathrm{DF}}$ as
\begin{align}
	\label{eq12}
	R_{\mathrm{DF}}^{\mathrm{up}}=\frac{1}{2}\log_2\left(1+N\min\left\{\frac{P_\mathrm{s}\|\mathbf{g}_1\|_1^2}{\sigma_\mathrm{r}^2},\frac{P_{\mathrm{r}}\|\mathbf{f}_2\|_1^2}{\sigma_\mathrm{d}^2}\right\}\right).
\end{align}
In contrast, for conventional FPAs, the performance bound in (16) is difficult to be reached due to the lack of DoFs to achieve the maximum channel power gains in (15). 

Next, given (16), we analyze its stochastic performance, i.e., the upper bound on the average achievable rate (AAR), denoted as $\mathbb{E}\left\{R_{\mathrm{DF}}\right\}$, with respect to $g_{1,\ell}\sim\mathcal{CN}(0,\rho_1^2/L_\mathrm{r}),1\leq\ell\leq L_{\mathrm{r}}$ and $f_{2,\ell}\sim\mathcal{CN}(0,\rho_2^2/L_\mathrm{t}),1\leq\ell\leq L_{\mathrm{t}}$.

\textit{\textbf{Proposition 1:} Under the ALR assumption, the AAR for the MA-enhanced DF relaying system is upper bounded by
	\begin{align}
		\label{eq13}
		\bar{R}_{\mathrm{DF}}^{\mathrm{up}}=\mathbb{E}\left\{R_{\mathrm{DF}}^{\mathrm{up}}\right\}\approx\frac{1}{2}\log_2\left(1+NI_1+NI_2\right),
	\end{align}
	where $I_1$ and $I_2$ are respectively given by
	\begin{subequations}
		\begin{align}
			\label{eq14}
			I_1&=\frac{1}{(L_\mathrm{t}-1)!}\sum_{k=0}^{L_\mathrm{r}-1}\frac{2\alpha_1^{L_\mathrm{t}+1}\alpha_2^{k+1}(k+L_\mathrm{t})!}{(\alpha_1+\alpha_2)^{k+L_\mathrm{t}+1}k!},\\
			I_2&=\frac{1}{(L_\mathrm{r}-1)!}\sum_{k=0}^{L_\mathrm{t}-1}\frac{2\alpha_1^{k+1}\alpha_2^{L_\mathrm{r}+1}(k+L_\mathrm{r})!}{(\alpha_1+\alpha_2)^{k+L_\mathrm{r}+1}k!},
		\end{align}
	\end{subequations}
	with $\alpha_1=\frac{\rho_1^2P_\mathrm{s}\left[(2L_\mathrm{r}-1)!!\right]^{1/L_\mathrm{r}}}{2L_\mathrm{r}^2\sigma_\mathrm{r}^2}$ and $\alpha_2=\frac{\rho_2^2P_\mathrm{r}\left[(2L_\mathrm{t}-1)!!\right]^{1/L_\mathrm{t}}}{2L_\mathrm{t}^2\sigma_\mathrm{d}^2}$}.

\textit{Proof: See Appendix A. $\hfill\square$}

Proposition 1 provides a closed-form expression for evaluating the stochastic performance of the MA-enhanced DF relaying system. However, the optimal MA positions are still yet to be determined. Besides, the above upper bounds are derived based on the ALR assumption. Hence, in the next subsection, we will develop a low-complexity algorithm to find the MA positions under the condition of a finite region size for antenna movement.

\subsection{Optimization Algorithm}
In this subsection, we propose an AO algorithm to solve (P1.1) and (P1.2) using the PGA method, as detailed below.

\textit{1) Optimization of $\{\mathbf{r}_n\}$:} Note that $\|\mathbf{h}_1(\tilde{\mathbf{r}})\|^2=|h_{1,1}(\mathbf{r}_1)|^2+|h_{1,2}(\mathbf{r}_2)|^2\cdots+|h_{1,N}(\mathbf{r}_N)|^2$, where each term is only related to a specific MA position. As a result, we propose to optimize each of the MAs' positions in an alternate manner. For the $n$-th MA's position, i.e., $\mathbf{r}_n,1\leq n \leq N$, the associated optimization problem can be expressed as
\begin{subequations}
	\begin{align}
		\max_{\mathbf{r}_n}& \quad |h_{1,n}(\mathbf{r}_n)|^2=\mathbf{f}_1^{\mathrm{H}}(\mathbf{r}_n)\mathbf{g}_1\mathbf{g}^{\mathrm{H}}\mathbf{f}_1(\mathbf{r}_n)\\
		\mathrm{s.t.}
		& ~\quad \mathbf{r}_n\in\mathcal{C}_\mathrm{r},\\
		& \quad \|\mathbf{r}_n-\mathbf{r}_m\|_2\geq D,~\forall m\neq n.
	\end{align}
\end{subequations}
By denoting the gradient of the objective function in (19a) with respect to $\mathbf{r}_n$ as $\nabla_{\mathbf{r}_{n}} |h_{1,n}(\mathbf{r}_{n})|^2$, the update rule for the $n$-th MA's position $\mathbf{r}_n$ during the $i$-th iteration of the PGA is given by
\begin{align}
	\mathbf{r}_{n}^{(i)}=\mathcal{B}\left\{\mathbf{r}_{n}^{(i-1)}+\eta^{(i)} \nabla_{\mathbf{r}_{n}} \left|h_{1,n}\left(\mathbf{r}_{n}^{(i-1)}\right)\right|^2\right\},
\end{align}
where $\mathcal{B}\{\mathbf{r}\}$ is a projection function to guarantee that each MA position does not exceed the finite region $\mathcal{C}_\mathrm{r}$, i.e.,
\begin{equation}	
	[\mathcal{B}\left(\mathbf{r}\right)]_p=
	\begin{cases}
		-A/2;&{\text{if}}~ [\mathbf{r}]_p<-A/2, \\
		[\mathbf{r}]_p;&{\text{if}}~ -A/2\leq[\mathbf{r}]_p\leq A/2, \\
		A/2;&{\text{if}}~ [\mathbf{r}]_p>A/2,
	\end{cases}
\end{equation}
where $[\cdot]_p$ denotes the $p$-th entry of the argument. Besides, in (20), $\eta^{(i)}\geq 0$ is the step size of the search along the gradient ascent direction in the $i$-th iteration. In particular, to guarantee that each $\mathbf{r}_{n}^{(i)}$ satisfies the minimum inter-MA distance constraint, we
should gradually decrease $\eta^{(i)}$ from a large positive initial value, denoted as $\eta^{(i)}=\eta_{\mathrm{ini}}$, by repeatedly shrinking it with a factor $\mu\in(0,1)$, i.e., $\eta^{(i)}\leftarrow \mu\eta^{(i)}$, until constraint (19c) is satisfied and $|h_{1,n}(\mathbf{r}_{n}^{(i)})|^2\geq|h_{1,n}(\mathbf{r}_{n}^{(i-1)})|^2$.  
Furthermore, by defining $\mathbf{G}=\mathbf{g}_1\mathbf{g}_1^{\mathrm{H}}$, the gradient of $|h_{1,n}(\mathbf{r}_n)|^2$ with respect to $\mathbf{r}_n$ can be derived in closed
form as:
\begin{subequations}
	\begin{align}
		\frac{\partial \left|h_{1,n}\left(\mathbf{r}_n\right)\right|^2}{\partial x_{r,n}}&=-\frac{2\pi}{\lambda}\sum_{\ell_1=1}^{L_r}\sum_{\ell_2=1}^{L_r}\left|\left[\mathbf{G}\right]_{\ell_1,\ell_2}\right|\notag\sin\left(\vartheta_{\ell_1,\ell_2}\left(\mathbf{r}_n\right)\right)\notag\\
		\times&(\sin\theta_{1,\ell_1}\cos\phi_{1,\ell_1}-\sin\theta_{1,\ell_2}\cos\phi_{1,\ell_2}),\\
		\frac{\partial \left|h_{1,n}\left(\mathbf{r}_n\right)\right|^2}{\partial y_{r,n}}&=-\frac{2\pi}{\lambda}\sum_{\ell_1=1}^{L_r}\sum_{\ell_2=1}^{L_r}\left|\left[\mathbf{G}\right]_{\ell_1,\ell_2}\right|\notag\sin\left(\vartheta_{\ell_1,\ell_2}\left(\mathbf{r}_n\right)\right)\notag\\
		\times&(\cos\theta_{1,\ell_1}-\cos\theta_{1,\ell_2}),
	\end{align}
\end{subequations}
with $\vartheta_{\ell_1,\ell_2}(\mathbf{r}_n)=\frac{2\pi}{\lambda}(\omega_{1,\ell_1}(\mathbf{r}_n)-\omega_{1,\ell_2}(\mathbf{r}_n))-\arg([\mathbf{G}]_{\ell_1,\ell_2})$. Note that if the step size $\eta^{(i)}$ is less than a small prescribed value $\eta_{\min}$, the PGA process for optimizing $\mathbf{r}_n$ can be declared as reaching its local maximum. In such case, the searching process for $\mathbf{r}_n$ should terminate accordingly. Otherwise, the iterations of the PGA for optimizing $\mathbf{r}_n$ will be repeated until a
maximum iteration number (denoted as $I_{\max}$) is reached.

\begin{algorithm}[t]
	\renewcommand{\algorithmicrequire}{\textit{~~Input:}}
	\renewcommand{\algorithmicensure}{\textit{~~Output:}}
	\caption{PGA-based algorithm for solving problem (19).}
	\begin{algorithmic}[1]
		\REQUIRE $L_{\mathrm{r}},\mathcal{C}_\mathrm{r},D,\mathbf{g}_1,\mathbf{r}_n,\{\theta_{1,\ell}\},\{\phi_{1,\ell}\},\eta_{\mathrm{ini}},\eta_{\min},I_{\max},\mu$\\
		\STATE Set the MA position as $\mathbf{r}_n^{(0)}=\mathbf{r}_n$.
		
		\FOR{$i = 1:1: I_{\max}$}
		\STATE Calculate the gradient $\nabla_{\mathbf{r}_{n}} \left|h_{1,n}\left(\mathbf{r}_{n}^{(i-1)}\right)\right|^2$ via (22).
		\STATE Initialize the step size as $\eta^{(i)}=\eta_{\mathrm{ini}}$.
		\STATE Update the MA position $\mathbf{r}_{n}^{(i)}$ via (20).
		\WHILE{$|h_{1,n}(\mathbf{r}_{n}^{(i)})|^2<|h_{1,n}(\mathbf{r}_{n}^{(i-1)})|^2$ or constraint (19c) is not satisfied}
		\STATE Shrink the step size $\eta^{(i)}\leftarrow \mu\eta^{(i)}$.
		\STATE Update the MA position $\mathbf{r}_{n}^{(i)}$ via (20).
		\ENDWHILE
		\IF{$\eta^{(i)}\leq\eta_{\min}$}
		\STATE Break.
		\ENDIF
		\ENDFOR
		\STATE Set the MA position as $\mathbf{r}_n=\mathbf{r}_n^{(i)}$.
		\RETURN $\mathbf{r}_n$.
	\end{algorithmic}
\end{algorithm}

The proposed PGA-based algorithm for solving problem (19) is summarized in Algorithm 1. The computational complexity of Algorithm 1 is analyzed
as follows. The computational complexity of calculating the gradient in line 3 is $\mathcal{O}\left(L_\mathrm{r}^2\right)$. The complexity of calculating the objective
function is $\mathcal{O}\left(L_\mathrm{r}\right)$, which entails the complexity of $\mathcal{O}\left(J_{\mathrm{max}}L_\mathrm{r}\right)$ in line 6-9, with $J_{\mathrm{max}}$ denoting the maximum number of iterations for searching the step size. Therefore, the overall computational complexity of Algorithm 1 for solving problem (19) is given by $\mathcal{O}\left(I_{\max}\left(L_\mathrm{r}^2+J_{\mathrm{max}}L_\mathrm{r}\right)\right)$.


\textit{2) Optimization of $\{\mathbf{t}_n\}$:} Similar to the optimization of $\mathbf{r}_n$, we can also expand the objective function in (P1.2) as $\|\mathbf{h}_2(\tilde{\mathbf{t}})\|^2=|h_{2,1}(\mathbf{t}_1)|^2+|h_{2,2}(\mathbf{t}_2)|^2\cdots+|h_{2,N}(\mathbf{t}_N)|^2$, where each term is only related to a specific MA position $\mathbf{t}_n$. As a result, we apply a similar AO algorithm to sequentially update the positions of the MAs. For the $n$-th MA's position in Stage II, the optimization problem can be expressed as
\begin{subequations}
	\begin{align}
		\quad\max_{\mathbf{t}_n}& \quad |h_{2,n}(\mathbf{t}_n)|^2=\mathbf{g}_2^{\mathrm{H}}(\mathbf{t}_n)\mathbf{f}_2\mathbf{f}_2^{\mathrm{H}}\mathbf{g}_2(\mathbf{t}_n)\\
		\mathrm{s.t.}
		& ~\quad \mathbf{t}_n\in\mathcal{C}_\mathrm{r},\\
		& \quad \|\mathbf{t}_n-\mathbf{t}_m\|_2\geq D,~\forall m\neq n.
	\end{align}
\end{subequations}
Since problem (23) has a similar structure as problem (19), we can simply modify Algorithm 1 to optimize the $n$-th MA's position at the relay in Stage II, i.e., $\mathbf{t}_n$, by replacing
\begin{align*}
	\left\{L_\mathrm{r},\mathbf{g}_1,\mathbf{r}_n,\left\{\mathbf{r}_m,m\neq n\right\}_{m=1}^N,\{\theta_{1,\ell}\}_{\ell=1}^{L_\mathrm{r}},\{\phi_{1,\ell}\}_{\ell=1}^{L_\mathrm{r}}\right\}
\end{align*}
therein with 
\begin{align*}
	\left\{L_\mathrm{t},\mathbf{f}_2,\mathbf{t}_n,\left\{\mathbf{t}_m,m\neq n\right\}_{m=1}^N,\{\theta_{2,\ell}\}_{\ell=1}^{L_\mathrm{t}},\{\phi_{2,\ell}\}_{\ell=1}^{L_\mathrm{t}}\right\}.
\end{align*} 
Besides, the corresponding computational complexity for optimizing $\mathbf{t}_n$ is $\mathcal{O}\left(I_{\max}\left(L_\mathrm{t}^2+J_{\mathrm{max}}L_\mathrm{t}\right)\right)$.

\begin{algorithm}[t]
	\renewcommand{\algorithmicrequire}{\textit{~~Input:}}
	\renewcommand{\algorithmicensure}{\textit{~~Output:}}
	\caption{Overall algorithm for solving (P1).}
	\begin{algorithmic}[1]
		\REQUIRE $N,L_{\mathrm{t}},L_{\mathrm{r}},\mathcal{C}_\mathrm{r},A,D,\mathbf{g}_1,\mathbf{f}_2,\{\theta_{1,\ell}\},
		\{\phi_{1,\ell}\},\{\theta_{2,\ell}\}$\\
		\qquad$\{\phi_{2,\ell}\},\eta_{\mathrm{ini}},\eta_{\min},I_{\max},\mu,\epsilon_1,\epsilon_2$\\
		\STATE Initialize a feasible solution $\{\mathbf{r}_n\}_{n=1}^N$ to (P1.1).
		\STATE Initialize a feasible solution $\{\mathbf{t}_n\}_{n=1}^N$ to (P1.2).
		
		\REPEAT
		\FOR{$n=1:1:N$}
		\STATE Given $\{\mathbf{r}_m,m\neq n\}_{m=1}^N$, solve problem (19) to update $\mathbf{r}_n$ by Algorithm 1.
		\ENDFOR
		\UNTIL{the increment in the objective of (13) is below $\epsilon_1$}
		
		\REPEAT
		\FOR{$n=1:1:N$}
		\STATE Given $\{\mathbf{t}_m,m\neq n\}_{m=1}^N$, solve problem (23) to update $\mathbf{t}_n$ similarly to Algorithm 1.
		\ENDFOR
		\UNTIL{the increment in the objective of (14) is below $\epsilon_2$}
		
		\RETURN $\{\mathbf{r}_n\}_{n=1}^N$ and $\{\mathbf{t}_n\}_{n=1}^N$.
	\end{algorithmic}
\end{algorithm}
\textit{3) Overall algorithm:} The overall algorithm for solving (P1) is summarized
in Algorithm 2. In lines 3-7, each MA position at the relay in Stage I is optimized alternately until the increment of $\|\mathbf{h}_1(\tilde{\mathbf{r}})\|^2$ is below a predefined threshold $\epsilon_1$. Similarly, in lines 8-12, each MA position at the relay in Stage II is optimized alternately until the increment of $\|\mathbf{h}_2(\tilde{\mathbf{t}})\|^2$ is below a predefined threshold $\epsilon_2$.
Since their objective functions are non-decreasing over the iterations and are upper bounded from the definition, the convergence of the AO iterations in Algorithm 2 can be guaranteed. 
Denote the maximum number of iterations for optimizing $\tilde{\mathbf{r}}$ and $\tilde{\mathbf{t}}$ as $T_{\max}^{(1)}$ and $T_{\max}^{(2)}$, respectively. Then, the computational complexity of Algorithm 2 is $\mathcal{O}(NI_{\max}T_{\max}^{(1)}(L_\mathrm{r}^2+J_{\mathrm{max}}L_\mathrm{r})+NI_{\max}T_{\max}^{(2)}(L_\mathrm{t}^2+J_{\mathrm{max}}L_\mathrm{t}))$.

\section{Performance Analysis and Proposed\\~Solution for AF Relaying}
In this section, we conduct performance analysis of the AF relaying system and solve (P2). To this end, we first equivalently transform (P2) into a more tractable form and reveal the structural properties of its optimal solution. Based on these properties, we derive an upper bound on $R_{\mathrm{AF}}$ under the ALR assumption, followed by a low-complexity algorithm proposed to obtain a high-quality suboptimal solution to (P2) based on its equivalent form.

\subsection{Performance Analysis}
\textit{1) Properties of the Optimal Solution to (P2):}
From (P2), it can be observed that the three high-dimensional  variables are highly coupled with each other in the objective function. To resolve this issue, we reveal the hidden separability of the optimization variables by introducing the following lemmas. 

\textit{\textbf{Lemma 2:} For any given $\tilde{\mathbf{r}}$ and $\tilde{\mathbf{t}}$, if (P2) is feasible, the optimal beamforming matrix, denoted as $\mathbf{W}^{\star}$, always satisfies $\mathrm{rank}(\mathbf{W}^{\star})=1$.}

\textit{Proof: See Appendix B. $\hfill\square$}

Based on Lemma 2, we can reduce the feasible region of (P2) by introducing a new constraint, i.e., $\mathrm{rank}(\mathbf{W})=1$, without losing its optimality. Therefore, by defining $\mathbf{W}=\gamma\mathbf{q}_2\mathbf{q}_1^\mathrm{H}$, where $\mathbf{q}_1\in\mathbb{C}^{N\times1}$ and $\mathbf{q}_1\in\mathbb{C}^{N\times1}$ are both unit-norm auxiliary variables, (P2) can be recast as
\begin{subequations}
	\begin{align}
		\text{(P2.1)}~\max_{\gamma,\mathbf{q}_1,\mathbf{q}_2,\tilde{\mathbf{r}},\tilde{\mathbf{t}}}& \quad \frac{P_\mathrm{s}\gamma^2\left|\mathbf{h}_2^\mathrm{H}(\tilde{\mathbf{t}})\mathbf{q}_2\right|^2\left|\mathbf{q}_1^\mathrm{H}\mathbf{h}_1(\tilde{\mathbf{r}})\right|^2}{\sigma_\mathrm{r}^2\gamma^2\left|\mathbf{h}_2^\mathrm{H}(\tilde{\mathbf{t}})\mathbf{q}_2\right|^2+\sigma_\mathrm{d}^2}\\
		\mathrm{s.t.}
		& \quad 
		P_\mathrm{s}\gamma^2\left|\mathbf{q}_1^\mathrm{H}\mathbf{h}_1(\tilde{\mathbf{r}})\right|^2+\sigma_\mathrm{r}^2\gamma^2\leq P_{\mathrm{r}},\\
		& \quad \left\|\mathbf{q}_1\right\|=1,\left\|\mathbf{q}_2\right\|=1,\\
		& \quad \text{(12c)-(12f)}.\notag
	\end{align}
\end{subequations}
Next, we present the following Lemma 3 to show the equivalence between the optimal MA positions for (P2.1) and those for (P1.1) and (P1.2), as well as their relationship with the optimal AF beamforming matrix in (P2).

\textit{\textbf{Lemma 3:} In (P2.1), the optimal MA positions, denoted as $\tilde{\mathbf{r}}^{\star}$ and $\tilde{\mathbf{t}}^{\star}$, are equivalent to those for (P1.1) and (P1.2), based on which the optimal beamforming matrix for (P2) can be constructed as
	\begin{align}
		\mathbf{W}^{\star}=\beta\mathbf{h}_2\left(\tilde{\mathbf{t}}^{\star}\right)\mathbf{h}_1^\mathrm{H}\left(\tilde{\mathbf{r}}^{\star}\right),
	\end{align}
	where $\beta$ is a scaling factor satisfying the AF relay's power constraint in (24b), given by
	\begin{align}
		\beta=\frac{1}{\sqrt{\|\mathbf{h}_1(\tilde{\mathbf{r}}^{\star})\|^2\|\mathbf{h}_2(\tilde{\mathbf{t}}^{\star})\|^2\left(P_\mathrm{s}\|\mathbf{h}_1(\tilde{\mathbf{r}}^{\star})\|^2+\sigma_\mathrm{r}^2\right)/P_\mathrm{r}}}.
	\end{align}}

\textit{Proof: See Appendix C. $\hfill\square$}

Hereto, it is found that although the three high-dimensional  variables $\tilde{\mathbf{r}}$, $\tilde{\mathbf{t}}$ and $\mathbf{W}$ are highly coupled with each other in (P2), their optimal solutions can be obtained separately by first maximizing the channel gains between the source/relay and the relay/destination, and then calculating the optimal beamforming matrix at the relay via (25).

\textit{2) Performance Analysis:}
 Recall that the maximum channel gains are upper bounded by (15). Therefore, by substituting (15) and (25) into (10), an upper bound on $R_{\mathrm{AF}}$ is derived as
\begin{align}
	&R_{\mathrm{AF}}^{\mathrm{up}}=\notag\\
	&\frac{1}{2}\log_2\left(1+\frac{N^2P_\mathrm{s}P_\mathrm{r}\|\mathbf{g}_1\|_1^2\|\mathbf{f}_2\|_1^2}{N\sigma_\mathrm{r}^2P_\mathrm{r}\|\mathbf{f}_2\|_1^2+N\sigma_{\mathrm{d}}^2P_\mathrm{s}\|\mathbf{g}_1\|_1^2+\sigma_{\mathrm{r}}^2\sigma_\mathrm{d}^2}\right).
\end{align}
To analyze the stochastic performance of the considered MA-enhanced AF relaying system, we present the following proposition which provides a closed-form upper bound on the AAR, i.e., $\mathbb{E}\left\{R_{\mathrm{AF}}\right\}$.

\textit{\textbf{Proposition 2:} Under the ALR assumption, the AAR for the MA-enhanced AF relaying system is upper bounded by
	\begin{align}
		\bar{R}_{\mathrm{AF}}^{\mathrm{up}}&=\mathbb{E}\left\{R_{\mathrm{AF}}^{\mathrm{up}}\right\}\notag\\
		&\approx\frac{1}{2}\log_2\left(1+\frac{N^2P_\mathrm{s}P_\mathrm{r}V_1V_2}{N\sigma_\mathrm{r}^2P_\mathrm{r}V_2+N\sigma_\mathrm{d}^2P_\mathrm{s}V_1+\sigma_{\mathrm{r}}^2\sigma_\mathrm{d}^2}\right),
	\end{align}
	with $V_1=\rho_1^2\left[1+\frac{(L_\mathrm{r}-1)\pi}{4}\right]$ and $V_2=\rho_2^2\left[1+\frac{(L_\mathrm{t}-1)\pi}{4}\right]$.}
	
	\textit{Proof:} See Appendix D. $\hfill\square$
	
	Proposition 2 provides a closed-form upper bound for evaluating the stochastic performance of the MA-enhanced AF relaying system. Similar to the DF relaying system, the upper bounds in (27) and (28) are both based on the ALR assumption. Thus, in the next subsection, we also develop a low-complexity algorithm to obtain a suboptimal solution under the condition of a finite region size for antenna movement.

\subsection{Optimization Algorithm}
\begin{algorithm}[t]
	\renewcommand{\algorithmicrequire}{\textit{~~Input:}}
	\renewcommand{\algorithmicensure}{\textit{~~Output:}}
	\caption{Overall algorithm for solving (P2).}
	\begin{algorithmic}[1]
		\REQUIRE $N,L_{\mathrm{t}},L_{\mathrm{r}},\mathcal{C}_\mathrm{r},A,D,\mathbf{g}_1,\mathbf{f}_2,\{\theta_{1,\ell}\},
		\{\phi_{1,\ell}\},\{\theta_{2,\ell}\}$\\
		\qquad$\{\phi_{2,\ell}\},\eta_{\mathrm{ini}},\eta_{\min},I_{\max},\mu,\epsilon_1,\epsilon_2$\\
		\STATE Initialize a feasible solution $\{\mathbf{r}_n\}_{n=1}^N$ to (P1.1).
		\STATE Initialize a feasible solution $\{\mathbf{t}_n\}_{n=1}^N$ to (P1.2).
		
		\REPEAT
		\FOR{$n=1:1:N$}
		\STATE Given $\{\mathbf{r}_m,m\neq n\}_{m=1}^N$, solve problem (19) to update $\mathbf{r}_n$ by Algorithm 1.
		\ENDFOR
		\UNTIL{the increment in the objective of (13) is below $\epsilon_1$}
		
		\REPEAT
		\FOR{$n=1:1:N$}
		\STATE Given $\{\mathbf{t}_m,m\neq n\}_{m=1}^N$, solve problem (23) to update $\mathbf{t}_n$ similarly to Algorithm 1.
		\ENDFOR
		\UNTIL{the increment in the objective of (14) is below $\epsilon_2$}
		
		\STATE Obtain $\mathbf{W}$ according to (25).
		
		\RETURN $\{\mathbf{r}_n\}_{n=1}^N$, $\{\mathbf{t}_n\}_{n=1}^N$ and $\mathbf{W}$.
	\end{algorithmic}
\end{algorithm}
According to Lemma 3, the optimal solution to (P2) can be obtained by first solving two independent subproblems, i.e.,  (P1.1) and (P1.2), and then calculating $\mathbf{W}$ via (25) with given $\tilde{\mathbf{r}}$ and $\tilde{\mathbf{t}}$. Note that (P1.1) and (P1.2) have already been discussed and solved in Section IV. Therefore, we can make a slight modification of Algorithm 2 to effectively solve (P2) by adding a new step to optimize the AF beamforming matrix at the relay,
as show in line 13 of Algorithm 3. Since the computational complexity of calculating $\mathbf{W}$ in line 13 is  $\mathcal{O}(N(L_{\mathrm{r}}+L_{\mathrm{t}}+N))$, the total complexity of Algorithm 3 is $\mathcal{O}(NI_{\max}T_{\max}^{(1)}(L_\mathrm{r}^2+J_{\mathrm{max}}L_\mathrm{r})+NI_{\max}T_{\max}^{(2)}(L_\mathrm{t}^2+J_{\mathrm{max}}L_\mathrm{t})+N(L_{\mathrm{r}}+L_{\mathrm{t}}+N))$.

\section{Numerical Results}
In this section, numerical results are presented to evaluate the
performance of our proposed algorithms for MA-enhanced relaying systems. 

\subsection{Simulation Setup and Benchmark Schemes}
In the simulation, the total numbers of receive and transmit channel paths
at the relay are respectively set as $L_{\mathrm{r}}=L_{\mathrm{t}}=L=5$. The elevation and azimuth angles of the corresponding AoAs and AoDs are randomly generated within $[-\pi/2,\pi/2]$. The minimum inter-MA distance is set as $D=\lambda/2$. The large-scale fading coefficients for the channels between the source/relay and the relay/destination are set as unity, i.e., $\rho_1^2=\rho_2^2=1$, such that the channel power gains are normalized over the noise power and then we can set $\sigma_{\mathrm{r}}^2=\sigma_{\mathrm{d}}^2=\sigma^2=1$. The transmit power of the source is set the same as the relay power budget, i.e., $P_{\mathrm{s}}=P_{\mathrm{r}}=P$, and we define the average receive/transmit SNR at the relay as $P/\sigma^2$. For
Algorithm 1, the maximum number of iterations is set as $I_{\mathrm{max}}=300$; the initial step size for backtracking line search is set as $\eta_{\mathrm{ini}}=10\lambda$; The factor of shrinking the
step size is set as $\mu=0.5$. For both Algorithms 2 and 3, the threshold for terminating the iteration is set as $\epsilon_1=\epsilon_2=10^{-3}$. All the results in this section are
averaged over $10^{3}$ Monte Carlo simulations.
Besides, we consider the following benchmark schemes for performance comparison.
\begin{itemize}
	\item \textbf{FPA}:  The relay is equipped with an FPA-based uniform planar array with $N$ antennas, where any two adjacent antennas are spaced by half wavelength, $\lambda/2$.
	
	\item \textbf{AS}:  The relay is equipped with an FPA-based uniform planar array with multiple antennas spaced by half wavelength, $\lambda/2$.  In each channel realization, $N$ antennas are selected via exhaustive search to
	maximize the achievable rate at the destination.
	
	\item \textbf{One-time position adjustment (OTPA)}: The relay is still equipped with $N$ MAs but the positions of them are adjusted only once to cater to both information reception and transmission in two stages. The associated antenna position optimization problem can be solved by applying a similar PGA-based algorithm, for which the details are omitted for brevity.
	
\end{itemize}

\subsection{MA-Enhanced DF Relaying System}
In this subsection, we consider the MA-enhanced DF relaying system and provide numerical results to validate the correctness of our analytical results and the effectiveness of our proposed algorithm in Section IV. 

\begin{figure}[!t]
	\centering
	\includegraphics[width=0.43\textwidth]{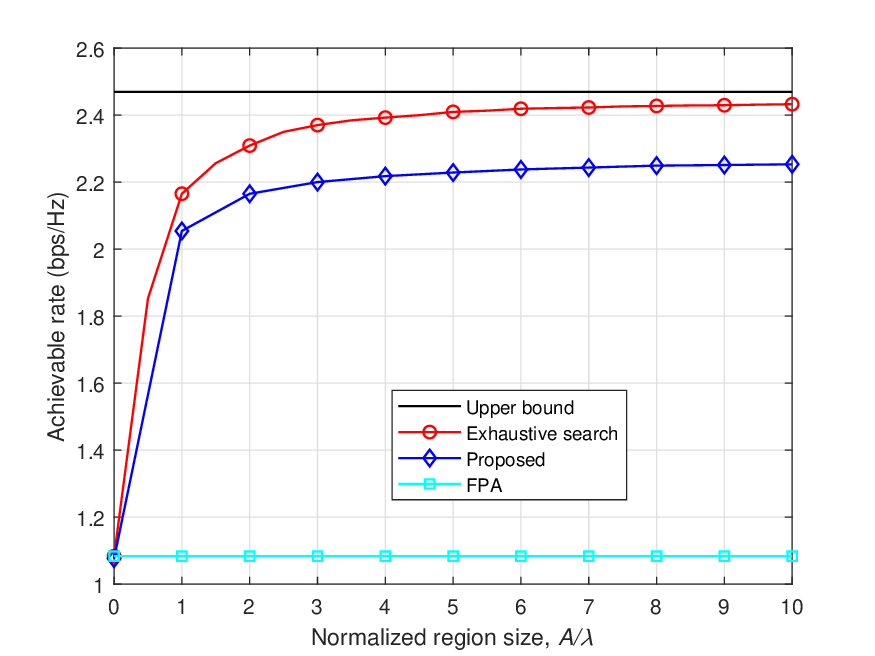}
	\caption{Achievable rates of the DF relaying system versus the
		normalized region size with $N=1$ and $\text{SNR}=10\text{~dB}$.}
	\label{single_antenna_DF_vs_A}
	\vspace{-0.25cm}
\end{figure}
\begin{figure}[!t]
	\centering
	\includegraphics[width=0.43\textwidth]{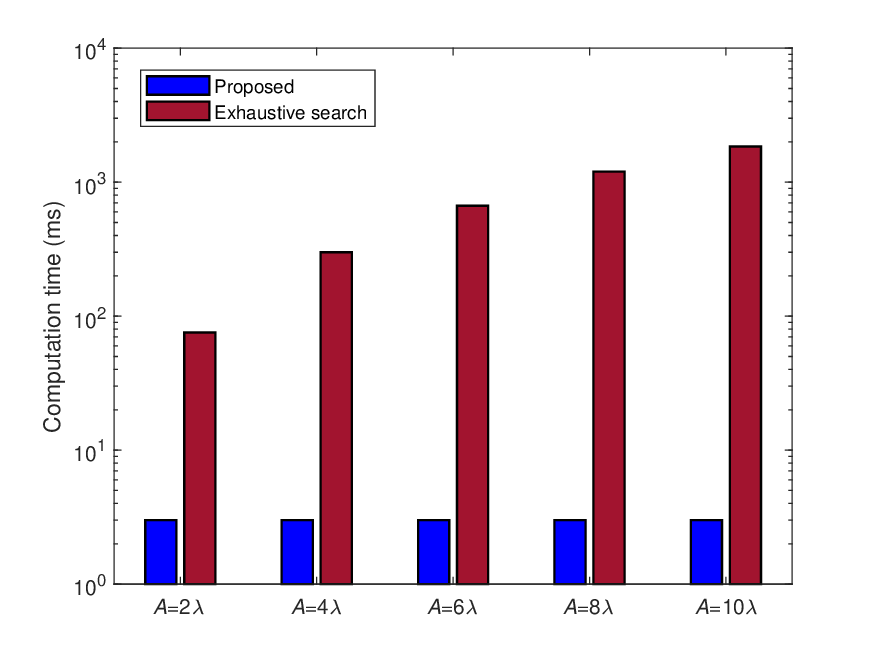}
	\caption{Computational time of our proposed algorithm and the exhaustive search with $N=1$.}
	\label{computation_time}
	\vspace{-0.25cm}
\end{figure}
\begin{figure}[!t]
	\centering
	\subfloat[Channel power gains between the source and the relay]{\includegraphics[width=0.45\textwidth]{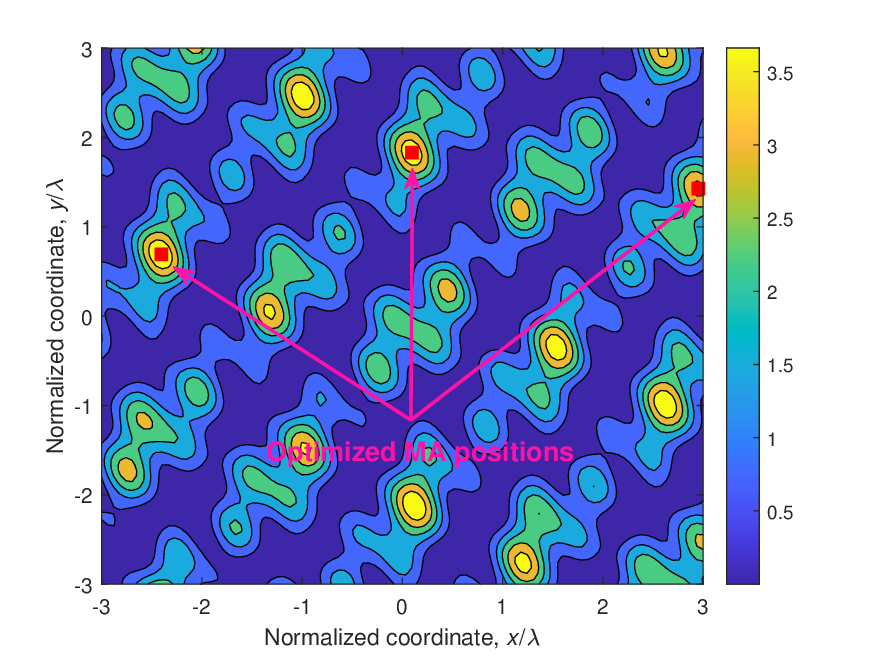}%
		\label{MA_positions_r}}
	\vspace{-0.25cm}
	\hfill
	\subfloat[Channel power gains between the relay and the destination]{\includegraphics[width=0.45\textwidth]{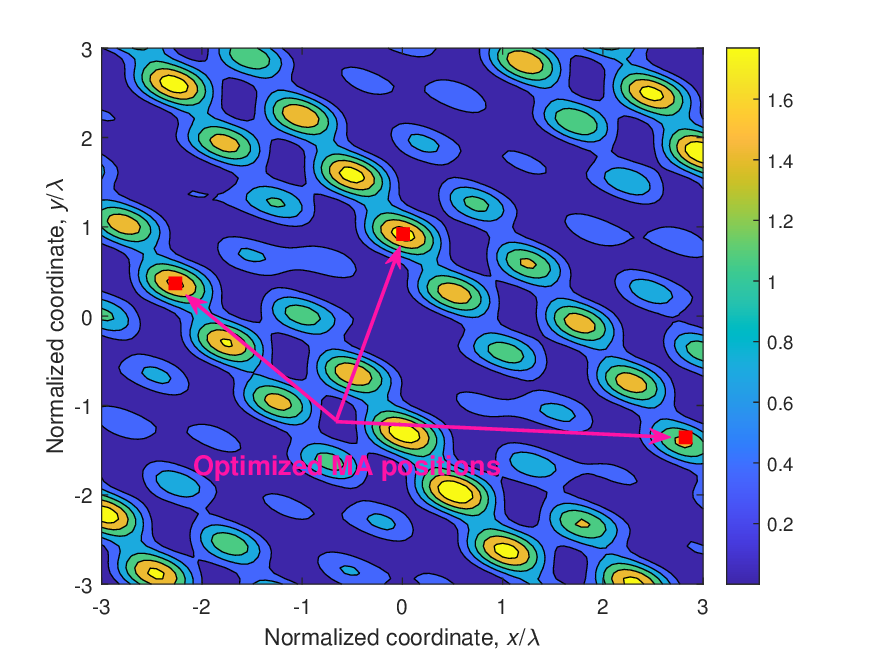}%
		\label{MA_positions_t}}
	\caption{Variation of the channel power gains within the normalized antenna moving region with $N=3$, $\text{SNR}=10\text{~dB}$, and $A=6\lambda$.}
	\label{MA_positions}
	\vspace{-0.25cm}
\end{figure}
First, in Fig. 3, we plot the achievable rates of different schemes versus the normalized region size for antenna movement at the relay under the single-MA setup, i.e., $N=1$, with $\text{SNR}=10\text{~dB}$. In order to validate the analytical upper bound presented in Proposition 1 and evaluate the optimality gap of our proposed algorithm, we also show the performance of the exhaustive search, which can obtain the optimal MA positions in both two stages. To this end, for each channel realization, we discretize the MA moving region at the relay into multiple discrete grids of equal size,
$\lambda/100\times\lambda/100$, and search among them to find the grid that yields the maximum channel power gain between the source/relay and the relay/destination. From Fig. 3, it can be observed that the achievable rates of our proposed algorithm and exhaustive search increase with the region size, since more available spatial diversity gain can be reaped from the antenna movement with a larger moving region. In contrast, the achievable rates achieved by the conventional FPA-based system, remain unchanged for different region sizes. This is because the antennas are located at fixed positions and thus do not have the flexibility to exploit the spatial variation of wireless channels for achieving more favorable channel conditions. Thanks to the spatial diversity, it is also observed that our proposed scheme can significantly outperform the traditional FPA-based system. In particular, our proposed algorithm can achieve a close performance to the exhaustive search (with the gap less than 0.2 bps/Hz) over the whole range, which implies that our proposed algorithm can achieve a near-optimal performance. It is also interesting to note that the achievable rates of the exhaustive search (averaged over 1000 channel realizations) approach the upper bound on the AAR provided in Proposition 1 under the ALR assumption, when the region size is is sufficiently large (larger than $7\lambda$). This phenomenon validates our analytical results in Proposition 1. 

In Fig. 4, we show the computational time of our proposed algorithm and the exhaustive search under the single-MA setup, i.e., $N=1$, with different sizes of the antenna moving region, i.e., $A=2\lambda,4\lambda,6\lambda,8\lambda$ and $10\lambda$. It can be observed that with any given size $A$, the computational time of our proposed algorithm remains nearly unchanged and is much shorter than that of the exhaustive search. On the contrary, the computational time of the exhaustive search increases rapidly as $A$ grows. In particular, as $A \geq 7\lambda$ as observed from Fig. 3 to achieve a close performance to the upper bound, the computational time of the exhaustive search is hundreds of times that of our proposed algorithm. As a result, for the more general multi-MA case, i.e., $N\geq2$, it would be more difficult to perform an exhaustive search over all possible MA position combinations due to the prohibitive computational complexity. For example, in the case of $N=3$ MAs in a $7\lambda\times7\lambda$ region, the total number of  MA position combinations for exhaustive search is $\bigg(\begin{aligned}
	&4900\\
	&~~3
\end{aligned}\bigg)=1.9592 \times 10^{10}$, which is unaffordable in practice. Therefore, in the subsequent simulation results with multi-MA setups, we will not include the exhaustive search for performance comparison.

\begin{figure}[!t]
	\centering
	\includegraphics[width=0.43\textwidth]{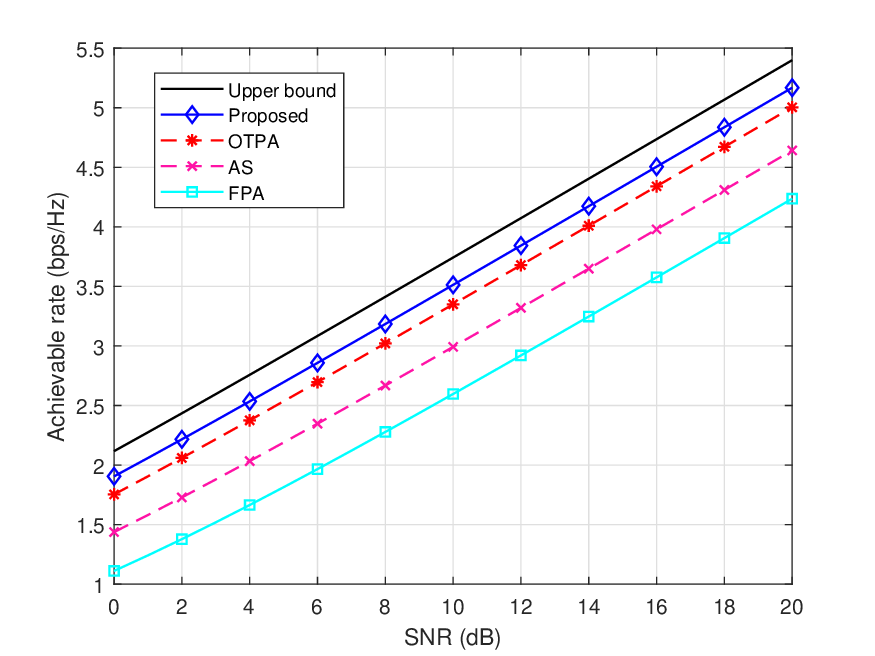}
	\caption{Achievable rates of the DF relaying system versus
		the SNR with $N=6$ and $A=10\lambda$.}
	\label{multi_antenna_DF_vs_SNR}
	\vspace{-0.25cm}
\end{figure}
To visualize the impact of MA position optimization, in Figs. 5(a) and 5(b), we demonstrate the variation of the source-relay and relay-destination channel power gains within the normalized antenna moving region, respectively, with $N=3$, $\text{SNR}=10~\text{dB}$, and $A=6\lambda$. From both Fig. 5(a) and Fig. 5(b), it can be observed that there are many local maxima of the channel power gain (in bright yellow color). These local maxima correspond to the positions with favorable channel conditions, which help improve the system performance. On the other hand, several deep-fading areas (in dark blue color) are also observed from Figs. 5(a) and 5(b), which are detrimental to the information transmission and reception. Fig. 5 shows that the optimized MA positions by our proposed algorithm are precisely located in the local maxima, which implies the efficacy of our proposed algorithm for improving both source-relay and relay-destination channel conditions, even within a small region size.
\begin{figure}[!t]
	\centering
	\includegraphics[width=0.43\textwidth]{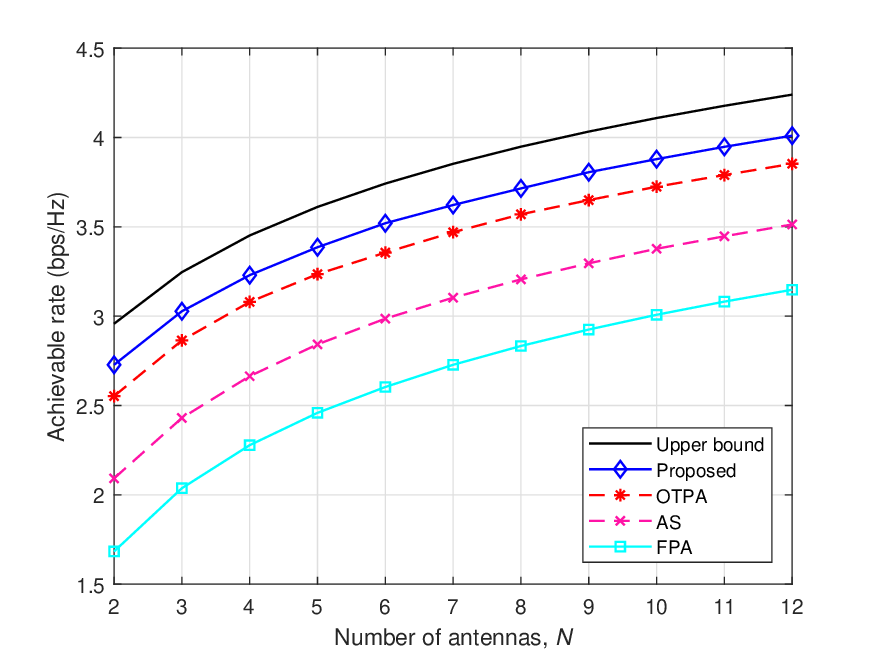}
	\caption{Achievable rates of the DF relaying system versus
		the number of antennas at the relay with $\text{SNR}=10\text{~dB}$ and $A=10\lambda$.}
	\label{multi_antenna_DF_vs_N}
	\vspace{-0.25cm}
\end{figure}

In Fig. 6, we show the achievable rates of our proposed algorithm and other benchmark schemes versus the average SNR (in dB). The number of MAs at the relay is set to $N=6$, and the size of their moving region is set to $A=10\lambda$. For the AS and FPA schemes, the relay is equipped with $2N$ and $N$ FPAs, respectively. As can be observed, the achievable rates of all considered schemes increase with the SNR. This is expected since a higher SNR should allow for more information to be transmitted. We also observe that our proposed solution can achieve a significant performance gain compared to the FPA and AS schemes over the whole range of SNR considered, by fully exploiting the spatial variation of wireless channels via local antenna movement. In particular, in the case of $\text{SNR}=10~\text{dB}$, our proposed solution can achieve $35.3\%$ and $17.4\%$ performance improvements over the FPA and AS schemes, respectively. Besides, there also exists a slight performance gap between our proposed scheme and the OTPA scheme. The reason is that the OTPA scheme needs to accommodate both information reception and transmission of the relay with only one-time antenna position adjustment, thus limiting the design flexibility. Nonetheless, some local maxima for source-relay and relay-destination channel power gains may be located closely, as observed from Fig. 5. Hence, the performance gap between the OTPA scheme and our proposed two-stage position adjustment scheme can be moderate, which also provides a more cost-effective solution to MA-aided DF relaying systems, especially in delay-sensitive scenarios..


Fig. \ref{multi_antenna_DF_vs_N} plots the achievable rates of different schemes versus the number of antennas at the relay, $N$. The average SNR is set to SNR $=10~\mathrm{dB}$ and the size of the antenna moving region is set to $A=10\lambda$. For the AS and FPA schemes, the relay is equipped with $2N$ and $N$ FPAs, respectively. It is observed that our proposed solution can achieve a close performance to the upper bound and a higher achievable rate compared to the FPA scheme and other benchmark schemes under different values of $N$. Specifically, for $N=9$, our proposed scheme can achieve $30.1\%,15.5\%,$ and $4.3\%$ performance gains over its FPA, AS, and OTPA counterparts, respectively. Besides, with the increase of antenna number, all considered schemes can achieve a
higher date rate thanks to the  improvement in spatial diversity gain and beamforming gain.

\subsection{MA-Enhanced AF Relaying System}
\begin{figure}[!t]
	\centering
	\includegraphics[width=0.43\textwidth]{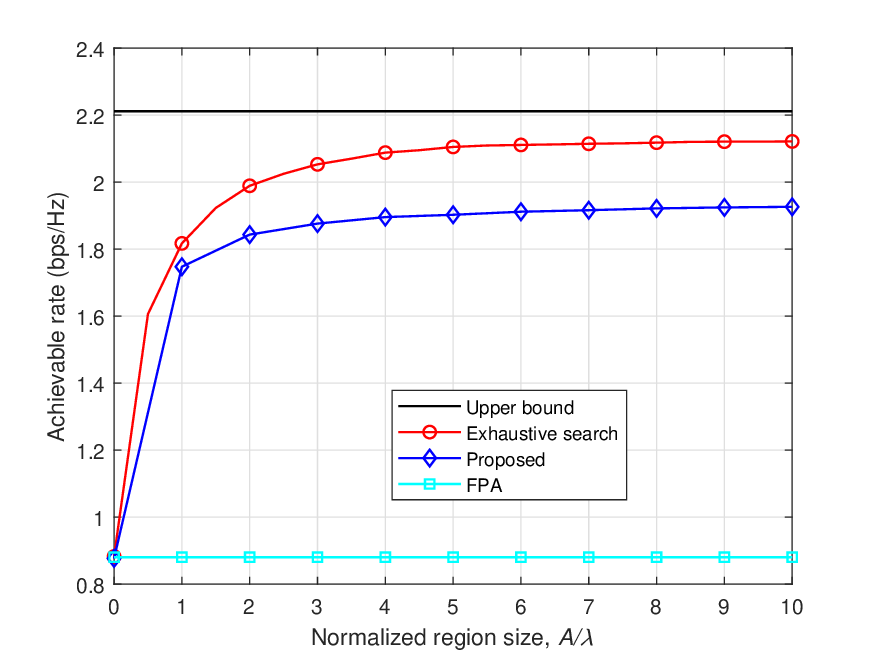}
	\caption{Achievable rates of the AF relaying system versus the
		normalized region size with $N=1$ and $\text{SNR}=10\text{~dB}$.}
	\label{single_antenna_AF_vs_A}
	\vspace{-0.25cm}
\end{figure}
Next, we consider the MA-enhanced AF relaying system and provide numerical results to validate the correctness of our analytical results and the effectiveness of our proposed algorithm in Section V. 

In Fig. \ref{single_antenna_AF_vs_A}, we show the achievable rates of our proposed solutions versus the normalized region size for antenna movement at the relay under the single-MA setup, i.e., $N=1$, with $\text{SNR}=10\text{~dB}$. Similar to the observations derived from Fig. 3, with the increasing region size, both of our proposed algorithm and the exhaustive search can achieve a higher data rate thanks to the enhanced spatial diversity gain. Moreover, our proposed algorithm can achieve a better performance compared to the conventional FPA-based scheme and a close performance to the exhaustive search (less than 0.2 bps/Hz). Meanwhile, we also observe that the achievable rate obtained by the exhaustive search approaches the performance upper bound as $A \ge 7\lambda$, which is consistent with our analytical results provided in Proposition 2.

\begin{figure}[!t]
	\centering
	\includegraphics[width=0.43\textwidth]{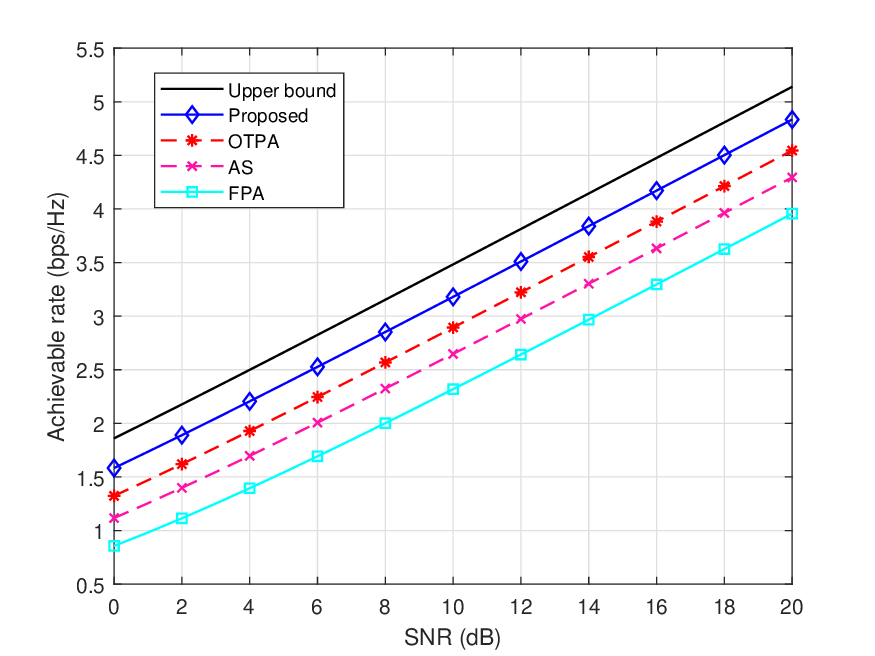}
	\caption{Achievable rates of the AF relaying system versus
		the SNR with $N=6$ and $A=10\lambda$.}
	\label{multi_antenna_AF_vs_SNR}
	\vspace{-0.25cm}
\end{figure}
\begin{figure}[!t]
	\centering
	\includegraphics[width=0.43\textwidth]{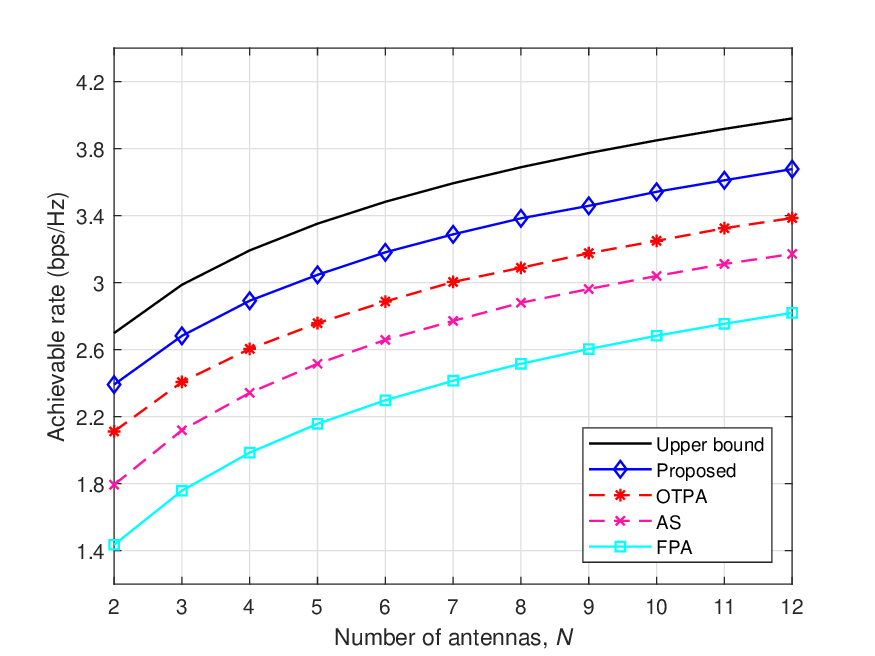}
	\caption{Achievable rates of the AF relaying system versus
		the number of antennas at the relay with $\text{SNR}=10\text{~dB}$ and $A=10\lambda$.}
	\label{multi_antenna_AF_vs_N}
	\vspace{-0.25cm}
\end{figure}
In Fig. \ref{multi_antenna_AF_vs_SNR}, we compare the achievable rates of our proposed scheme and other benchmark schemes with varying SNR (in dB). It is also observed that our proposed solution can achieve the highest data rate among all considered schemes with any given SNR. In particular, as $\text{SNR}=10\text{~dB}$, our proposed scheme can achieve $37.1\%,20.1\%$ and $9.8\%$ performance improvements over the FPA, AS, and OTPA schemes, respectively. Besides, there also exists a moderate performance gap between our proposed scheme and the OTPA scheme. 

Finally, Fig. \ref{multi_antenna_AF_vs_N} shows the achievable rates by different schemes versus the number of antennas at the relay, $N$. The average SNR at the relay is set to $\text{SNR}=10\text{~dB}$ and the size of the antenna moving region at the relay is set to $A=10\lambda$. It is observed again that our proposed scheme can achieve a  better performance compared to the FPA scheme and other benchmark schemes under different values of $N$. Specifically, for $N=9$, our proposed scheme can achieve $28.5\%,13.0\%$ and $5.3\%$ performance improvements over its FPA, AS, and OTPA counterparts, respectively.

\section{Conclusion}
In this paper, we conducted performance analysis and optimization for both MA-enhanced DF and AF relaying systems. Different from most related works requiring only one-time position adjustment, the MA positions in our considered relaying systems need to be adjusted in two stages for information reception and transmission, respectively. For the MA-enhanced DF relaying system, we derived a closed-form expression of the upper bound on the achievable rate at the destination under the ALR assumption, following which a low-complexity algorithm based on PGA and AO was proposed to find a suboptimal solution for the  MA positions in two stages. While for the MA-enhanced AF relaying system, we derived its performance upper bound in closed form and solved the associated antenna position and AF beamforming optimization problem by showing their separable property at the optimality. 
Simulation results validated our analytical upper bounds  and the effectiveness of our proposed algorithms for both MA-enhanced DF and AF relaying systems. It was also shown that our proposed scheme can significantly outperform the conventional FPA-based scheme and achieve a small performance gap with the analytical upper bound. Future works could be carried out towards more general system setups with a multi-antenna source, a multi-antenna destination, and/or multiple relays.

{\appendices
	\section{Proof of Proposition 1}
	First, we define $X^2=\max_{\mathbf{r}\in\mathcal{C}_\mathrm{r}} P_\mathrm{s}\|\mathbf{h}_1(\tilde{\mathbf{r}})\|^2/\sigma_\mathrm{r}^2$ and $Y^2=\max_{\mathbf{r}\in\mathcal{C}_\mathrm{r}} P_{\mathrm{r}}\|\mathbf{h}_2(\tilde{\mathbf{t}})\|^2/\sigma_\mathrm{d}^2$. Then, the upper bound on $\mathbb{E}\{R_{\mathrm{DF}}\}$ can be equivalently expressed as
	\begin{align}
		\bar{R}_{\mathrm{DF}}^{\mathrm{up}}=\mathbb{E}\left[\frac{1}{2}\log_2\left(1+\min\left\{X^2,Y^2\right\}\right)\right].
	\end{align}
	Note that the exact expression of (29) is difficult to derive. To resolve this challenge, based on the conclusion in \cite{ref38}, we can obtain a tight approximation of $\bar{R}_{\mathrm{DF}}^{\mathrm{up}}$, given by
	\begin{align}
		\label{eq29}
		\bar{R}_{\mathrm{DF}}^{\mathrm{up}}\approx\frac{1}{2}\log_2\left(1+\mathbb{E}\left[\min\left\{X^2,Y^2\right\}\right]\right).
	\end{align}
	Besides, based on Lemma 1, we have
	\begin{align}
		X^2\triangleq N\tilde{X}^2\approx\frac{NP_\mathrm{s}}{\sigma_\mathrm{r}^2}\left(\sum_{\ell=1}^{L_\mathrm{r}}\left|g_{1,\ell}\right|\right)^2,\\Y^2\triangleq N\tilde{Y}^2\approx\frac{NP_{\mathrm{r}}}{\sigma_\mathrm{d}^2}\left(\sum_{\ell=1}^{L_\mathrm{r}}\left|f_{2,\ell}\right|\right)^2.
	\end{align}
	Since $g_{1,\ell}$ and $f_{2,\ell}$ are i.i.d. CSCG random variables with zero mean and variance $\rho_1^2/L_\mathrm{r}$ and $\rho_2^2/L_\mathrm{t}$, $|g_{1,\ell}|$ and $|f_{2,\ell}|$ are i.i.d. Rayleigh-distributed random variables with parameters of $\rho_1^2/(2L_\mathrm{r})$ and $\rho_2^2/(2L_\mathrm{t})$, respectively. Thus, $\tilde{X}$ and $\tilde{Y}$ are the sum of $L_\mathrm{r}$ and $L_\mathrm{t}$ i.i.d. Rayleigh-distributed variables, respectively, whose cumulative distribution function (CDF) and probability density function (PDF) are \cite{ref37}
	\begin{align}
		F_{\tilde{X}}(\zeta)&=1-e^{-\frac{\zeta^2}{2\alpha_1}}\sum_{k=0}^{L_\mathrm{r}-1}\frac{(\zeta^2/2\alpha_1)^k}{k!},\\
		f_{\tilde{X}}(\zeta)&=\frac{\zeta^{2L_\mathrm{r}-1}e^{-\frac{\zeta^2}{2\alpha_1}}}{2^{L_\mathrm{r}-1}\alpha_1^{L_\mathrm{r}}(L_\mathrm{r}-1)!},
	\end{align}
	and
	\begin{align}
		F_{\tilde{Y}}(\zeta)&=1-e^{-\frac{\zeta^2}{2\alpha_2}}\sum_{k=0}^{L_\mathrm{t}-1}\frac{(\zeta^2/2\alpha_2)^k}{k!}\\
		f_{\tilde{Y}}(\zeta)&=\frac{\zeta^{2L_\mathrm{t}-1}e^{-\frac{\zeta^2}{2\alpha_2}}}{2^{L_\mathrm{t}-1}\alpha_2^{L_\mathrm{t}}(L_\mathrm{t}-1)!}.
	\end{align}
	Subsequently, by defining $Z=\min\{\tilde{X},\tilde{Y}\}$,  we can re-express $\mathbb{E}[\min\{X^2,Y^2\}]$ as
	\begin{align}
		\mathbb{E}\left[\min\left\{X^2,Y^2\right\}\right]&=\mathbb{E}\left[NZ^2\right]=N\int_{z=0}^{+\infty}z^2f_Z(z)\mathrm{d}z\notag\\
		&=N\underbrace{\int_{z=0}^{+\infty}z^2\left(1-F_X(z)\right)f_Y(z)\mathrm{d}z}_{I_1}\notag\\
		&+N\underbrace{\int_{z=0}^{+\infty}\left(1-F_Y(z)\right)f_X(z)\mathrm{d}z}_{I_2}.
	\end{align}
	Substituting (33) and (36) into $I_1$, we can obtain (18a), i.e.,
	\begin{align}
		I_1&=\int_{z=0}^{+\infty}\frac{z^{2L_\mathrm{t}+1}e^{-\left(\frac{1}{2\alpha_1}+\frac{1}{2\alpha_2}\right)z^2}\times\sum_{k=0}^{L_\mathrm{r}-1}\frac{(z^2/2\alpha_1)^k}{k!}\mathrm{d}z}{2^{L_\mathrm{t}-1}\alpha_2^{L_\mathrm{t}}(L_\mathrm{t}-1)!}\notag\\
		&=\frac{\sum_{k=0}^{L_\mathrm{r}-1}\frac{1}{(2\alpha_1)^kk!}\int_{z=0}^{+\infty}z^{2k+2L_\mathrm{t}+1}e^{-\left(\frac{1}{2\alpha_1}+\frac{1}{2\alpha_2}\right)z^2}\mathrm{d}z}{2^{L_\mathrm{t}-1}\alpha_2^{L_\mathrm{t}}(L_\mathrm{t}-1)!}\notag\\
		&=\frac{\sum_{k=0}^{L_\mathrm{r}-1}\frac{1}{(2\alpha_1)^kk!}\int_{u=0}^{+\infty}u^{k+L_\mathrm{t}}e^{-\left(\frac{1}{2\alpha_1}+\frac{1}{2\alpha_2}\right)u}\mathrm{d}u}{2^{L_\mathrm{t}}\alpha_2^{L_\mathrm{t}}(L_\mathrm{t}-1)!}\notag\\
		&=\frac{1}{(L_\mathrm{t}-1)!}\sum_{k=0}^{L_\mathrm{r}-1}\frac{2\alpha_1^{L_\mathrm{t}+1}\alpha_2^{k+1}\Gamma(k+L_\mathrm{t}+1)}{(\alpha_1+\alpha_2)^{k+L_\mathrm{t}+1}k!}\notag\\
		&=\frac{1}{(L_\mathrm{t}-1)!}\sum_{k=0}^{L_\mathrm{r}-1}\frac{2\alpha_1^{L_\mathrm{t}+1}\alpha_2^{k+1}(k+L_\mathrm{t})!}{(\alpha_1+\alpha_2)^{k+L_\mathrm{t}+1}k!},
	\end{align} 
	with the values of the definite integral $\int_{u=0}^{+\infty}u^{n-1}e^{-\lambda u}\mathrm{d}u=\frac{\Gamma(n)}{\lambda^n}$, where $\Gamma(n)$ is the Gamma function defined as $\Gamma(n)=(n-1)!$. Then, by substituting (34) and (35) into $I_2$, we can obtain (18b) in the same way.
	Eventually, we can obtain (17) by substituting (37) into (30).
	
	\section{Proof of Lemma 2}
	For any given $\tilde{\mathbf{r}}$ and $\tilde{\mathbf{t}}$, by applying the property of Kronecker product\cite{ref30}, i.e.,
	\begin{align}
		&\mathrm{vec}(\mathbf{A}_1\mathbf{A}_2\mathbf{A}_3)=\left(\mathbf{A}_3^\mathrm{T}\otimes\mathbf{A}_1\right)\cdot\mathrm{vec}(\mathbf{A}_2),\\
		&\left(\mathbf{A}_1\otimes\mathbf{A}_2\right)^\mathrm{T}=\mathbf{A}_1^\mathrm{T}\otimes\mathbf{A}_2^\mathrm{T},
	\end{align}
	we can re-express (P2) as
	\begin{subequations}
		\begin{align}
			\max_{\mathbf{w}}& \quad \frac{P_\mathrm{s}\mathbf{w}^\mathrm{H}\mathbf{h}\mathbf{h}^\mathrm{H}\mathbf{w}}{\sigma_\mathrm{r}^2\mathbf{w}^\mathrm{H}\mathbf{A}\mathbf{A}^\mathrm{H}\mathbf{w}+\sigma_\mathrm{d}^2}\\
			\mathrm{s.t.}
			& \quad 
			\mathbf{w}^\mathrm{H}\left(P_\mathrm{s}\mathbf{B}\mathbf{B}^\mathrm{H}+\sigma_\mathrm{r}^2\mathbf{I}\right)\mathbf{w}\leq P_{\mathrm{r}},
		\end{align}
	\end{subequations}
	where $\mathbf{w}=\mathrm{vec}(\mathbf{W}),\mathbf{h}=\mathbf{h}_1^*\otimes\mathbf{h}_2,\mathbf{A}=\mathbf{I}\otimes\mathbf{h}_2$, and $\mathbf{B}=\mathbf{h}_1^*\otimes\mathbf{I}$. In problem (41), it can be easily proved that constraint (41b) is active at the optimum. Otherwise, we can always multiply $\mathbf{w}$ by a scalar $a~(a>1)$ to increase the value of the objective function. 
	As a result, problem (41) can be equivalently transformed into
	\begin{align}
		\max_{\mathbf{w}} \quad \frac{\mathbf{w}^\mathrm{H}\mathbf{X}_1\mathbf{w}}{\mathbf{w}^\mathrm{H}\mathbf{X}_3\mathbf{w}} \quad
		\mathrm{s.t.}
		~
		\mathbf{w}^\mathrm{H}\mathbf{X}_2\mathbf{w}=P_{\mathrm{r}},
	\end{align}
	where $\mathbf{X}_1=P_\mathrm{s}\mathbf{h}\mathbf{h}^\mathrm{H},\mathbf{X}_2=P_\mathrm{s}\mathbf{B}\mathbf{B}^\mathrm{H}+\sigma_\mathrm{r}^2\mathbf{I}$, and $\mathbf{X}_3=\sigma_\mathrm{r}^2\mathbf{A}\mathbf{A}^\mathrm{H}+\sigma_\mathrm{d}^2P_{\mathrm{r}}^{-1}\mathbf{X}_2$. Since the objective function of problem (42) is a generalized Rayleigh quotient, its optimal solution can be given by
	\begin{align}
		\mathbf{w}^{\star}=\xi\mathbf{X}_3^{-1}\mathbf{h},
	\end{align}
	where $\xi$ is a scalar satisfying the transmit power constraint at the relay, i.e., 
	\begin{align}
		\xi=\sqrt{P_\mathrm{r}}\left(\left\|\mathbf{X}_2^{\frac{1}{2}}\mathbf{X}_3^{-1}\mathbf{h}\right\|\right)^{-1}.
	\end{align}
	Then, from the definition of Kronecker product, we can expand the expression of $\mathbf{h}$ as
	\begin{align}
		\mathbf{h}=\left[h_{1,1}^*\mathbf{h}_2^\mathrm{T},\cdots,h_{1,N}^*\mathbf{h}_2^\mathrm{T}\right]^\mathrm{T},
	\end{align}
	where all components $(h_{1,n}^*\mathbf{h}_2^\mathrm{T},1\leq n\leq N)$ are $1\times N$ vectors linearly dependent with each other. As a result, the optimal solution to problem (42) can be further expanded as
	\begin{align}
		\mathbf{w}^{\star}=\begin{bmatrix}
			\xi h_{1,1}^*\mathbf{X}_3^{-1}\mathbf{h}_2\\
			\vdots\\
			\xi h_{1,N}^*\mathbf{X}_3^{-1}\mathbf{h}_2
		\end{bmatrix}
	\end{align}
	where $\xi h_{1,n}^*\mathbf{X}_3^{-1}\mathbf{h}_2,1\leq n \leq N$, are also linearly dependent with each other. Finally, the optimal solution for $\mathbf{W}$ can be constructed as
	\begin{align}
		\mathbf{W}^{\star}=\mathrm{vec}^{-1}\left(\mathbf{w}^{\star}\right)=\left[\xi h_{1,1}^*\mathbf{X}_3^{-1}\mathbf{h}_2,\cdots,\xi h_{1,N}^*\mathbf{X}_3^{-1}\mathbf{h}_2\right].
	\end{align}
	From (47), it is easy to show that $\mathbf{W}^{\star}$ is a rank-one matrix, since all columns therein are linearly dependent with each other and not equal to zero. Thus, the proof is complete.
	
	\section{Proof of Lemma 3}
	It is easy to show that constraint (24b) is always active at its optimum, i.e.,
	\begin{align}
		P_\mathrm{s}\gamma^2\left|\mathbf{q}_1^\mathrm{H}\mathbf{h}_1(\tilde{\mathbf{r}})\right|^2+\sigma_\mathrm{r}^2\gamma^2=P_{\mathrm{r}}.
	\end{align}
	Therefore, the objective function in (24a) can be re-written as
		\begin{align}
			S= \frac{P_\mathrm{s}P_\mathrm{r}\left|\mathbf{h}_2^\mathrm{H}(\tilde{\mathbf{t}})\mathbf{q}_2\right|^2\left|\mathbf{h}_1^\mathrm{H}(\tilde{\mathbf{r}})\mathbf{q}_1\right|^2}{P_\mathrm{r}\sigma_{\mathrm{r}}^2\left|\mathbf{h}_2^\mathrm{H}(\tilde{\mathbf{t}})\mathbf{q}_2\right|^2+P_\mathrm{s}\sigma_{\mathrm{d}}^2\left|\mathbf{h}_1^\mathrm{H}(\tilde{\mathbf{r}})\mathbf{q}_1\right|^2+\sigma_{\mathrm{r}}^2\sigma_{\mathrm{d}}^2},
		\end{align}
	which is a monotonically increasing function with respect to both $\left|\mathbf{h}_1^\mathrm{H}(\tilde{\mathbf{r}})\mathbf{q}_1\right|^2$ and $\left|\mathbf{h}_2^\mathrm{H}(\tilde{\mathbf{t}})\mathbf{q}_2\right|^2$. As a result, the optimal $\mathbf{q}_1^{\star}$ and $\mathbf{q}_2^{\star}$ can be respectively chosen as
	\begin{align}
		\mathbf{q}_1^{\star}&=\mathbf{h}_1(\tilde{\mathbf{r}})/\|\mathbf{h}_1(\tilde{\mathbf{r}})\|,\\
		\mathbf{q}_2^{\star}&=\mathbf{h}_2(\tilde{\mathbf{t}})/\|\mathbf{h}_2(\tilde{\mathbf{t}})\|.
	\end{align}
	Then, by substituting (50) and (51) into (49), the optimization problem can be re-formulated as
		\begin{align}
			\max_{\tilde{\mathbf{r}},\tilde{\mathbf{t}}}& \quad \frac{P_\mathrm{s}P_\mathrm{r}\left\|\mathbf{h}_2^\mathrm{H}(\tilde{\mathbf{t}})\right\|^2\left\|\mathbf{h}_1^\mathrm{H}(\tilde{\mathbf{r}})\right\|^2}{P_\mathrm{r}\sigma_{\mathrm{r}}^2\left\|\mathbf{h}_2^\mathrm{H}(\tilde{\mathbf{t}})\right\|^2+P_\mathrm{s}\sigma_{\mathrm{d}}^2\left\|\mathbf{h}_1^\mathrm{H}(\tilde{\mathbf{r}})\right\|^2+\sigma_{\mathrm{r}}^2\sigma_{\mathrm{d}}^2}\\
			\mathrm{s.t.}
			& \quad \text{(12c)-(12f)}.\notag
		\end{align}
		Note that $\left\|\mathbf{h}_1^\mathrm{H}(\tilde{\mathbf{r}})\right\|^2$ is only related to $\tilde{\mathbf{r}}$, and $\left\|\mathbf{h}_2^\mathrm{H}(\tilde{\mathbf{t}})\right\|^2$ is only related to $\tilde{\mathbf{t}}$. Besides, the objective in (52) is a monotonically increasing function with respect to both $\left\|\mathbf{h}_1^\mathrm{H}(\tilde{\mathbf{r}})\right\|^2$ and $\left\|\mathbf{h}_2^\mathrm{H}(\tilde{\mathbf{t}})\right\|^2$. Therefore, the optimal solution for problem (52), i.e., $\tilde{\mathbf{r}}^{\star}$ and $\tilde{\mathbf{t}}^{\star}$, are equivalent to those for  (P1.1) and (P1.2), such that the channel gains between the source/relay and the relay/destination can be maximized independently.
	 
	 	Finally, with given $\left\{\tilde{\mathbf{r}}^{\star},\tilde{\mathbf{t}}^{\star}.\mathbf{q}_1^{\star},\mathbf{q}_2^{\star}\right\}$, we can obtain the optimal $\gamma^{\star}$ from (48), as $\gamma^{\star}=\sqrt{P_\mathrm{r}/(P_\mathrm{s}\|\mathbf{h}_1(\tilde{\mathbf{r}}^{\star})\|^2+\sigma_{\mathrm{r}}^2)}.$
	 	Then, by substituting $\gamma^{\star}$, $\mathbf{q}_1^{\star}$ and $\mathbf{q}_2^{\star}$ into $\mathbf{W}$, we can eventually derive the optimal beamforming matrix $\mathbf{W}^{\star}$ in (25).
	    Based on the above, the proof is complete.
	    
	    \section{Proof of Proposition 2}
	    According to the conclusion in \cite[Lemma 1]{ref38}, $\bar{R}_{\mathrm{AF}}^{\mathrm{up}}$ can be approximated as
	    \begin{align}
	    	\bar{R}_{\mathrm{AF}}^{\mathrm{up}}\approx\frac{1}{2}\log_2\left(1+\frac{N^2P_\mathrm{s}P_\mathrm{r}V_1V_2}{N\sigma_\mathrm{r}^2P_\mathrm{r}V_2+N\sigma_\mathrm{d}^2P_\mathrm{s}V_1+\sigma_\mathrm{r}^2\sigma_\mathrm{d}^2}\right)
	    \end{align}
	    where $V_1=\mathbb{E}\left\{\|\mathbf{g}_1\|_1^2\right\}$ and $V_2=\mathbb{E}\left\{\|\mathbf{f}_2\|_1^2\right\}$. Since $g_{1,\ell},1\leq \ell \leq L_{\mathrm{r}},$ are i.i.d. CSCG random variables with mean $0$ and variance $\rho_1^2/L_{\mathrm{r}}$, $|g_{1,\ell}|,1\leq \ell \leq L_\mathrm{r},$ are i.i.d. Rayleigh distributed variables with their PDFs given by
	    \begin{align}
	    	f_{|g_{1,\ell}|}(\zeta)=\frac{2L_\mathrm{r}\zeta}{\rho_1^2} e^{\frac{-L_\mathrm{r}\zeta^2}{\rho_1^2}},~1\leq \ell \leq L_\mathrm{r}.
	    \end{align}
	    As a result, we have
	    \begin{align}
	    	V_1=&\mathbb{E}\left\{\|\mathbf{g}_1\|_1^2\right\}\notag\\
	    	=&\int_{0}^{\infty}\cdots\int_{0}^{\infty}\left(\sum_{i=1}^{L_\mathrm{r}}\zeta_i\right)^2\times\prod_{i=1}^{L_\mathrm{r}}\frac{2L_\mathrm{r}\zeta_i}{\rho_1^2} e^{\frac{-L_\mathrm{r}\zeta_i^2}{\rho_1^2}}\mathrm{d}\zeta_1\cdots\mathrm{d}\zeta_{L_\mathrm{r}}\notag\\
	    	=&\sum_{i=1}^{L_\mathrm{r}}\sum_{j \neq i}^{L_\mathrm{r}}
	    	\int_{0}^{\infty}\frac{2L_\mathrm{r}
	    	\zeta_i^2}{\rho_1^2}e^{\frac{-L_\mathrm{r}\zeta_i^2}{\rho_1^2}}\mathrm{d}\zeta_i\times\int_{0}^{\infty}\frac{2L_\mathrm{r}\zeta_j^2}{\rho_1^2}e^{\frac{-L_\mathrm{r}\zeta_j^2}{\rho_1^2}}\mathrm{d}\zeta_j\notag\\
	    	&+ \sum_{i=1}^{L_\mathrm{r}}\int_{0}^{\infty}\frac{2L_\mathrm{r}\zeta_i^3}{\rho_1^2}e^{\frac{-L_\mathrm{r}\zeta_i^2}{\rho_1^2}}\mathrm{d}\zeta_i\notag\\
	    	=&\rho_1^2\left[1+\frac{(L_\mathrm{r}-1)\pi}{4}\right].
	    \end{align}
	    with the values of the definite integrals $\int_{0}^{\infty}\frac{2a
	    \zeta^2}{\rho_1^2}e^{\frac{-a\zeta^2}{\rho_1^2 }}\mathrm{d}\zeta=\sqrt{\frac{\pi}{4a}}\rho_1$ and $\int_{0}^{\infty}\frac{2a\zeta^3}{\rho_1^2}e^{\frac{-a\zeta^2}{\rho_1^2}}\mathrm{d}\zeta=\frac{\rho_1^2}{a}$. Similarly, we can obtain
	    \begin{align}
	    	V_2&=\mathbb{E}\left\{\|\mathbf{f}_2\|_1^2\right\}=\rho_2^2\left[1+\frac{(L_\mathrm{t}-1)\pi}{4}\right].
	    \end{align}
	    Finally, (28) can be obtained by substituting (55) and (56) into (53) and the proof is thus complete.
	
}

\bibliography{reference}
\bibliographystyle{IEEEtran}

\vfill
\end{document}